\documentclass[english,aps,notitlepage,nofootinbib,tightenlines,11pt]{revtex4-1}
\usepackage[T1]{fontenc}
\usepackage[latin9]{inputenc}
\usepackage{color}
\usepackage{amsmath}
\usepackage{graphicx}
\usepackage{esint}

\makeatletter
\pdfoutput=1
\usepackage[colorlinks=true,citecolor=blue,linkcolor=blue]{hyperref}
\usepackage{afterpage}
\usepackage{slashed}
\usepackage{extarrows}

\allowdisplaybreaks

\makeatother

\usepackage{babel}
\begin{document}

\title{The $\nu_{R}$-philic scalar: its loop-induced interactions \\
and Yukawa forces in LIGO observations}

\author{Xun-Jie Xu}

\affiliation{\textcolor{black}{Max-Planck-Institut f\"ur Kernphysik, Postfach
103980, D-69029 Heidelberg, Germany}.}

\date{\today}
\begin{abstract}
Right-handed neutrinos ($\nu_{R}$) are often considered as a portal
to new hidden physics. It is tempting to consider a gauge singlet
scalar $(\phi)$ that exclusively couples to $\nu_{R}$ via a $\nu_{R}\nu_{R}\phi$
term.  Such a $\nu_{R}$-philic scalar does not interact with charged
fermions at tree level but loop-induced effective interactions are
inevitable, which are systematically investigated in this work. 
The magnitude of the loop-induced  couplings coincidentally meets
 the current sensitivity of fifth-force searches. In particular,
the loop-induced coupling to muons  could be tested in the recent
LIGO observations of neutron star mergers as there might be a sizable
Yukawa force in the binary system  mediated by the $\nu_{R}$-philic
scalar. 
\end{abstract}
\maketitle

\section{Introduction}

Right-handed neutrinos ($\nu_{R}$) are one of the most intriguing
pieces to be added to the Standard Model (SM). Not only can they resolve
several problems of the SM including neutrinos masses, dark mater,
and baryon asymmetry of the universe,\footnote{See, e.g., the so-called $\nu$MSM~\cite{Asaka:2005pn,Asaka:2005an}
which extends the SM by $\nu_{R}$ to incorporate neutrino masses,
dark mater, and leptogenesis simultaneously.} their singlet nature under the SM gauge symmetry  also allows for
couplings to hidden or dark sectors, a feature known as the neutrino
portal to physics beyond the SM. 

Among various new physics extensions built on $\nu_{R}$, a gauge
singlet scalar $\phi$ coupled exclusively to $\nu_{R}$, referred
to as the $\nu_{R}$-philic scalar, is arguably the simplest.\footnote{It has recently been shown that the $\nu_{R}$-philic scalar could
assist low-scale leptogenesis~\cite{Alanne:2018brf}.} At tree level, the $\nu_{R}$-philic scalar does not interact directly
with normal matter that consists of electrons and quarks, which implies
that it might have been well hidden from low-energy laboratory searches.
At the one-loop level, there are loop-induced couplings of $\phi$
to charged fermions, which are  suppressed by neutrino masses ($m_{\nu}$)
in the framework of Type I seesaw~\cite{Minkowski:1977sc,yanagida1979proceedings,GellMann:1980vs, glashow1979future,mohapatra1980neutrino}.
The suppression can be understood from that in the zero limit of neutrino
masses, which corresponds to vanishing couplings of the SM Higgs to
$\nu_{R}$ and left-handed neutrinos ($\nu_{L}$), the $\nu_{R}$
sector would be entirely decoupled from the SM content. As we will
show, for electrons, the loop-induced effective Yukawa coupling is
of the order of
\begin{equation}
\frac{G_{F}m_{e}m_{\nu}}{16\pi^{2}}\sim{\cal O}\left(10^{-21}\right),\label{eq:m-83}
\end{equation}
where $G_{F}$ is the Fermi constant and $m_{e}$ is the electron
mass.

Despite the small value of the loop-induced coupling, the magnitude
coincides with the sensitivity of current precision tests of gravity.
For long-range forces mediated by ultra-light bosons coupled to electrons
or quarks, experimental tests of the strong (based on the lunar laser-ranging
technology~\cite{Turyshev:2006gm}) and weak (e.g., torsion-balance
experiments~\cite{Schlamminger:2007ht,Heckel:2008hw})  equivalence
principles are sensitive to  Yukawa/gauge couplings spanning from
$10^{-20}$ to $10^{-24}$. Very recently, gravitational waves from
black hole (BH) and neutron star (NS) binary mergers have been detected
by the LIGO/VIRGO collaboration~\cite{TheLIGOScientific:2017qsa,Abbott:2020khf},
providing novel methods to test theories of gravity as well as other
long-range forces~\cite{Croon:2017zcu,Baryakhtar:2017ngi,Sagunski:2017nzb,Hook:2017psm,Huang:2018pbu, Kopp:2018jom,Alexander:2018qzg,Choi:2018axi,Fabbrichesi:2019ema,Seymour:2019tir,Dror:2019uea}.
For instance, the process of BH superradiance can be used to exclude
a wide range of ultra-light boson masses~\cite{Baryakhtar:2017ngi}.
The  sizable abundance of muons in NS binary systems allows us to
probe muonic forces as they could modify the orbital dynamics. It
is expected that~\cite{Dror:2019uea} current and future observations
of NS binaries are sensitive to muonic Yukawa/gauge couplings ranging
from $10^{-18}$ to $10^{-22}$ which, again, coincidentally covers
the theoretical expectation  of the loop-induced coupling for muons,
$G_{F}m_{\mu}m_{\nu}/(16\pi^{2})\sim10^{-19}$. 

In light of the frontiers of precision and novel tests of gravity
and gravity-like forces, it is important to perform an in-depth study
on the loop-induced interactions of the $\nu_{R}$-philic scalar,
which is the main goal of this work. We note here that in the seminal
work on majorons~\cite{Chikashige:1980ui}, similar loop-induced interactions
have been computed and confronted with experimental limits in the
1980s. More recently, Ref.~\cite{Garcia-Cely:2017oco} studied majoron
decay caused by the loop-induced couplings to charged fermions. In
addition, majoron decay to photons is also possible at two-loop level~\cite{Heeck:2019guh}.
While the majoron considered in Refs.~\cite{Chikashige:1980ui, Garcia-Cely:2017oco,Heeck:2019guh}
is a pseudo-scalar boson, in this work we compute loop-induced interactions
for a generic scalar and take three lepton flavors into account, with
loop calculation details presented. The loop-induced interactions
computed in this work could be of importance in phenomenological studies
of long-range forces~\cite{Joshipura:2003jh,Grifols:2003gy,Bandyopadhyay:2006uh, GonzalezGarcia:2006vp, Nelson:2007yq,GonzalezGarcia:2008wk,Samanta:2010zh,Heeck:2010pg, Davoudiasl:2011sz, Heeck:2014zfa,      Chatterjee:2015gta, Bustamante:2018mzu,Khatun:2018lzs,Wise:2018rnb, Krnjaic:2017zlz,Berlin:2016woy,Brdar:2017kbt,Smirnov:2019cae,Babu:2019iml}.

The paper is organized as follows. In Sec.~\ref{sec:basic}, we briefly
review the Type I seesaw extended by a gauge singlet scalar, and derive
the tree-level interactions for later use. In Sec.~\ref{sec:Loop},
we compute the loop-induced interactions of $\phi$ with charged fermions.
The calculation, for simplicity, is first performed assuming only
one generation of leptons and then generalized to three flavors in
Sec.~\ref{sec:3nu-1}. In Sec.~\ref{sec:Phenomenology}, we confront
the theoretical predictions to experimental limits including searches
for long-range forces of normal matter and the LIGO observations of
NS events which are sensitive to muonic couplings. We conclude in
Sec.~\ref{sec:Conclusion} and delegate some details of our calculations
to the appendix.

\section{The model\label{sec:basic}}

\subsection{Notations\label{sub:Notations}}

Throughout this paper, Weyl spinors are frequently used in our discussions
for simplicity. On the other hand, for  Feynman diagram calculations,
Dirac or Majorana spinors are more convenient due to a variety of
techniques and especially many modern computation packages that have
been developed. As both will used in this paper, it is necessary to
clarify our notations regarding Weyl spinors versus Dirac/Majorana
spinors.

All four-component Dirac/Majorana spinors in this paper are denoted
by $\psi_{X}$ with some interpretative subscripts $X$. Otherwise,
they are Weyl spinors. For instance, $\nu_{L}$ and $\ell_{R}$ are
Weyl spinors of a left-handed neutrino  and a right-handed charged
lepton, respectively. In contrast to that, $\psi_{\ell}$ is a Dirac
spinor of a charged lepton containing both left- and right-handed
components.

For Weyl spinors, our notation follows the convention in Ref.~\cite{Dreiner:2008tw}.
For example,  the mass and kinetic terms of $\nu_{R}$ are 
\begin{equation}
M_{R}\nu_{R}\nu_{R}\equiv M_{R}\left(\nu_{R}\right)^{\alpha}\left(\nu_{R}\right)_{\alpha},\ \ \nu_{R}^{\dagger}\overline{\sigma}^{\mu}i\partial_{\mu}\nu_{R}\equiv\left(\nu_{R}^{\dagger}\right)_{\dot{\alpha}}\left(\overline{\sigma}^{\mu}\right)^{\dot{\alpha}\beta}i\partial_{\mu}\left(\nu_{R}\right)_{\beta}.\label{eq:m-13}
\end{equation}
Here and henceforth, the Weyl spinor indices $\alpha$, $\dot{\alpha}$,
$\beta$ will be suppressed. 

Dirac and Majorana spinors can be built from Weyl spinors. Hence the
 Dirac spinors of charged leptons and neutrinos can be written as
\begin{equation}
\psi_{\ell}=\left(\begin{array}{c}
\ell_{L}\\[2mm]
\ell_{R}^{\dagger}
\end{array}\right),\ \psi_{\nu}=\left(\begin{array}{c}
\nu_{L}\\[2mm]
\nu_{R}^{\dagger}
\end{array}\right).\label{eq:m-14}
\end{equation}
 The Majorana spinor of a neutrino mass eigenstate $\nu_{i}$ (where
$i=1$, 2, 3, $\cdots$) is defined as
\begin{equation}
\psi_{i}\equiv\left(\begin{array}{c}
\nu_{i}\\[2mm]
\nu_{i}^{\dagger}
\end{array}\right).\label{eq:m-15}
\end{equation}
Note that it is self-conjugate: $\psi_{i}^{c}=\psi_{i}$.   For
later convenience, some identities are listed below to convert Weyl
spinors into Dirac/Majorana spinors :
\begin{equation}
\nu_{i}\nu_{j}=\nu_{j}\nu_{i}=\overline{\psi_{i}}P_{L}\psi_{j},\ \ \nu_{i}^{\dagger}\nu_{j}^{\dagger}=\nu_{j}^{\dagger}\nu_{i}^{\dagger}=\overline{\psi_{i}}P_{R}\psi_{j},\ \ \nu_{i}^{\dagger}\overline{\sigma}^{\mu}\nu_{j}=\overline{\psi_{i}}\gamma^{\mu}P_{L}\psi_{j},\label{eq:m-18}
\end{equation}
\begin{equation}
\ell_{L}\nu_{i}=\nu_{i}\ell_{L}=\overline{\psi_{i}}P_{L}\psi_{\ell},\ \ \ell_{R}\nu_{i}=\nu_{i}\ell_{R}=\overline{\psi_{\ell}}P_{L}\psi_{i},\ \ \ell_{L}^{\dagger}\overline{\sigma}^{\mu}\nu_{i}=\overline{\psi_{\ell}}\gamma^{\mu}P_{L}\psi_{i},\label{eq:m-19}
\end{equation}
where $P_{L/R}\equiv(1\mp\gamma^{5})/2$ and $\gamma_{L}^{\mu}\equiv\gamma^{\mu}P_{L}$.

\subsection{Lagrangian}

We consider the SM extended by several right-handed neutrinos $\nu_{R}$
and a singlet scalar $\phi$. In Type I seesaw, the number of $\nu_{R}$
needs to be $\geq2$  in order to accommodate the observed neutrino
oscillation data. Let us start with one generation of leptons and
ignore the flavor structure (for the realistic case including three
generations, see Sec.~\ref{sec:3nu-1}).  The Lagrangian of $\nu_{R}$
and $\phi$ reads:
\begin{equation}
{\cal L}\supset\nu_{R}^{\dagger}\overline{\sigma}^{\mu}i\partial_{\mu}\nu_{R}+\frac{1}{2}(\partial\phi)^{2}+\frac{1}{2}m_{\phi}^{2}\phi^{2}+\left[\frac{M_{R}}{2}\thinspace\nu_{R}\nu_{R}+\frac{y_{R}}{2}\thinspace\nu_{R}\nu_{R}\phi+{\rm h.c.}\right].\label{eq:m}
\end{equation}
Here we assume $\phi$ is a real scalar or pseudo-scalar field. If
it is a complex field, one can decompose it as $\phi=\phi_{r}+i\phi_{i}$
with $\phi_{r}$ and $\phi_{i}$ being real scalar and pseudo-scalar
fields respectively. To make our calculation applicable to both scalar
and pseudo-scalar cases, we allow $y_{R}$ to be a complex coupling.

The Dirac masses of leptons are generated by 
\begin{equation}
{\cal L}\supset y_{\nu}\widetilde{H}^{\dagger}L\nu_{R}+y_{\ell}H^{\dagger}L\ell_{R}+{\rm h.c.},\label{eq:m-1}
\end{equation}
where $H$ is the SM Higgs doublet ($\widetilde{H}\equiv i\sigma_{2}H^{*}$),
$L=(\nu_{L},\ \ell_{L})^{T}$ is a left-handed lepton doublet, and
$\ell_{R}$ is a right-handed charged lepton. After electroweak symmetry
breaking, $\langle H\rangle=(0,\ v)^{T}/\sqrt{2}$, Eq.~(\ref{eq:m-1})
leads to the following mass terms:
\begin{equation}
{\cal L}\supset m_{D}\nu_{L}\nu_{R}+m_{\ell}\ell_{L}\ell_{R}+{\rm h.c.},\label{eq:m-2}
\end{equation}
where
\begin{equation}
m_{D}\equiv y_{\nu}\frac{v}{\sqrt{2}},\ m_{\ell}\equiv y_{\ell}\frac{v}{\sqrt{2}}.\label{eq:m-3}
\end{equation}
The Dirac and Majorana mass terms of neutrinos can be formulated as
\begin{equation}
{\cal L}_{\nu\thinspace{\rm mass}}=\frac{1}{2}(\nu_{L},\ \nu_{R})\left(\begin{array}{cc}
0 & m_{D}\\
m_{D} & M_{R}
\end{array}\right)\left(\begin{array}{c}
\nu_{L}\\
\nu_{R}
\end{array}\right),\label{eq:m-4}
\end{equation}
which then can be diagonalized by
\begin{equation}
\left(\begin{array}{c}
\nu_{L}\\
\nu_{R}
\end{array}\right)=U\left(\begin{array}{c}
\nu_{1}\\
\nu_{4}
\end{array}\right),\ U^{T}\left(\begin{array}{cc}
0 & m_{D}\\
m_{D} & M_{R}
\end{array}\right)U=\left(\begin{array}{cc}
m_{1}\\
 & m_{4}
\end{array}\right).\label{eq:m-5}
\end{equation}
Here $\nu_{1}$ and $\nu_{4}$ are the light and heavy mass eigenstates
with their masses determined by
\begin{equation}
m_{1}=\frac{1}{2}\left(\sqrt{4m_{D}^{2}+M_{R}^{2}}-M_{R}\right),\ m_{4}=\frac{1}{2}\left(\sqrt{4m_{D}^{2}+M_{R}^{2}}+M_{R}\right).\label{eq:m-17}
\end{equation}
The unitary matrix $U$ is parametrized as
\begin{equation}
U=\left(\begin{array}{cc}
-i\thinspace c_{\theta} & s_{\theta}\\
i\thinspace s_{\theta} & c_{\theta}
\end{array}\right),\label{eq:m-6}
\end{equation}
where $c_{\theta}\equiv\cos\theta$, $s_{\theta}\equiv\sin\theta$,
and 
\begin{equation}
\theta=\arctan\sqrt{m_{1}/m_{4}}.\label{eq:m-21}
\end{equation}
Eq.~(\ref{eq:m-6}) has been parametrized in such a way that $m_{D}$,
$M_{R}$, $m_{1}$ and $m_{4}$ are all positive numbers.

\subsection{Interactions in the mass basis}

Since $\nu_{L}$ and $\nu_{R}$ are not mass eigenstates,  we need
to reformulate neutrino interactions in the mass basis, i.e., the
basis of $\nu_{1}$ and $\nu_{4}$.  The two bases are related by
\begin{eqnarray}
\nu_{L} & = & -i\thinspace c_{\theta}\thinspace\nu_{1}+s_{\theta}\thinspace\nu_{4}\thinspace,\label{eq:m-8}\\
\nu_{R} & = & i\thinspace s_{\theta}\thinspace\nu_{1}+c_{\theta}\thinspace\nu_{4}\thinspace.\label{eq:m-9}
\end{eqnarray}

Neutrino interactions in the original basis (chiral basis) include
gauge interactions and Yukawa interactions, summarized as follows:
\begin{equation}
{\cal L}\supset\frac{g}{2c_{W}}Z_{\mu}\nu_{L}^{\dagger}\overline{\sigma}^{\mu}\nu_{L}+\left[\frac{g}{\sqrt{2}}W_{\mu}^{-}\ell_{L}^{\dagger}\overline{\sigma}^{\mu}\nu_{L}-y_{\nu}H^{+}\ell_{L}\nu_{R}+y_{\ell}H^{-}\nu_{L}\ell_{R}+\frac{y_{R}}{2}\thinspace\nu_{R}\nu_{R}\phi+{\rm h.c.}\right],\label{eq:m-10}
\end{equation}
where $g$ is the gauge coupling of $SU(2)_{L}$ in the SM, $c_{W}$
is the cosine of the Weinberg angle, and $H^{\pm}$ is the charged
component of $H$, i.e. the Goldstone boson associated to $W^{\pm}$.

Now applying the basis transformation in Eqs.~(\ref{eq:m-8}) and
(\ref{eq:m-9}) to Eq.~(\ref{eq:m-10}), we get
\begin{equation}
{\cal L}\supset g_{Z}^{ij}Z_{\mu}\nu_{i}^{\dagger}\overline{\sigma}^{\mu}\nu_{j}+\left[g_{W}^{i}W_{\mu}^{-}\ell_{L}^{\dagger}\overline{\sigma}^{\mu}\nu_{i}-y_{\nu}^{i}H^{+}\ell_{L}\nu_{i}+y_{\ell}^{i}H^{-}\nu_{i}\ell_{R}+\frac{y_{R}^{ij}}{2}\thinspace\nu_{i}\nu_{j}\phi+{\rm h.c.}\right].\label{eq:m-10-2}
\end{equation}
Here $i$ and $j$ take either 1 or 4. The couplings $g_{Z}^{ij}$,
$g_{W}^{i}$, $y_{\nu}^{i}$, $y_{\ell}^{i}$, $y_{R}^{ij}$ are given
by the following matrices or vectors:
\begin{eqnarray}
g_{Z}^{ij} & = & \frac{g}{2c_{W}}\left(\begin{array}{cc}
c_{\theta}^{2} & ic_{\theta}s_{\theta}\\
-ic_{\theta}s_{\theta} & s_{\theta}^{2}
\end{array}\right),\ \ y_{R}^{ij}=y_{R}\left(\begin{array}{cc}
-s_{\theta}^{2} & ic_{\theta}s_{\theta}\\
ic_{\theta}s_{\theta} & c_{\theta}^{2}
\end{array}\right),\label{eq:m-11}\\
g_{W}^{i} & = & \frac{g}{\sqrt{2}}(-ic_{\theta},\ s_{\theta}),\ \ y_{\nu}^{i}=y_{\nu}(is_{\theta},\ c_{\theta}),\ \ y_{\ell}^{i}=y_{\ell}(-ic_{\theta},\ s_{\theta}).\label{eq:m-12}
\end{eqnarray}
Eq.~(\ref{eq:m-10-2}) can be straightforwardly expressed in terms
of Dirac and Majorana spinors according to Eqs.~(\ref{eq:m-18})
and (\ref{eq:m-19}):
\begin{equation}
{\cal L}\supset g_{Z}^{ij}Z_{\mu}\overline{\psi_{i}}\gamma_{L}^{\mu}\psi_{j}+\left[g_{W}^{i}W_{\mu}^{-}\overline{\psi_{\ell}}\gamma_{L}^{\mu}\psi_{i}+H^{-}\overline{\psi_{\ell}}(y_{\ell}^{i}P_{L}-y_{\nu}^{i*}P_{R})\psi_{i}+\frac{y_{R}^{ij}}{2}\thinspace\overline{\psi_{i}}P_{L}\psi_{j}\phi+{\rm h.c.}\right].\label{eq:m-20}
\end{equation}
Note that in the mass basis, $\phi$ couples to both heavy and light
neutrinos but the coupling of the latter is suppressed by $s_{\theta}$.

\section{Loop-induced interactions of $\phi$ with charged leptons\label{sec:Loop}}

\begin{figure}[t]
\centering

\includegraphics[width=0.8\textwidth]{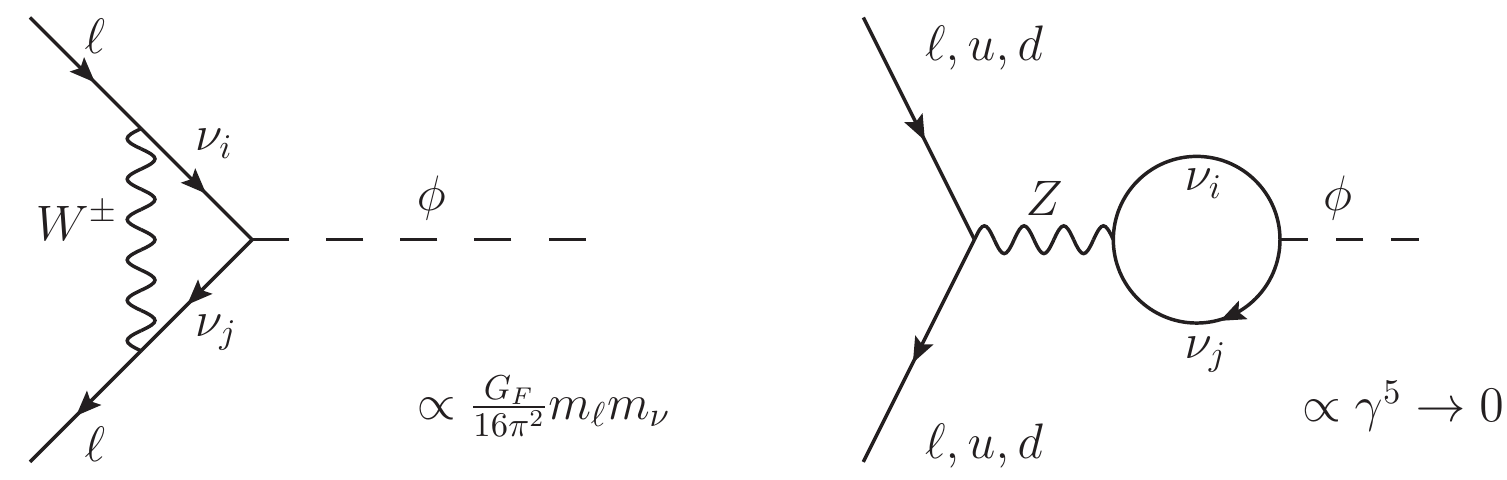}

\caption{One-loop diagrams that give rise to effective couplings of $\phi$
with charged leptons ($\ell$) or quarks ($u$, $d$). The left diagram
is computed in Eqs.~(\ref{eq:m-29})-(\ref{eq:F2}), and the right
diagram leads to a pseudo-scalar coupling (with $\gamma^{5}$), the
effect of which however is suppressed in unpolarized matter.  The
diagrams are presented in the mass basis ($\nu_{i}$ and $\nu_{j}$
are mass eigenstates). For an equivalent description in the chiral
basis, see Fig.~\ref{fig:W-chiral}. \label{fig:WZ}}

\includegraphics[width=0.5\textwidth]{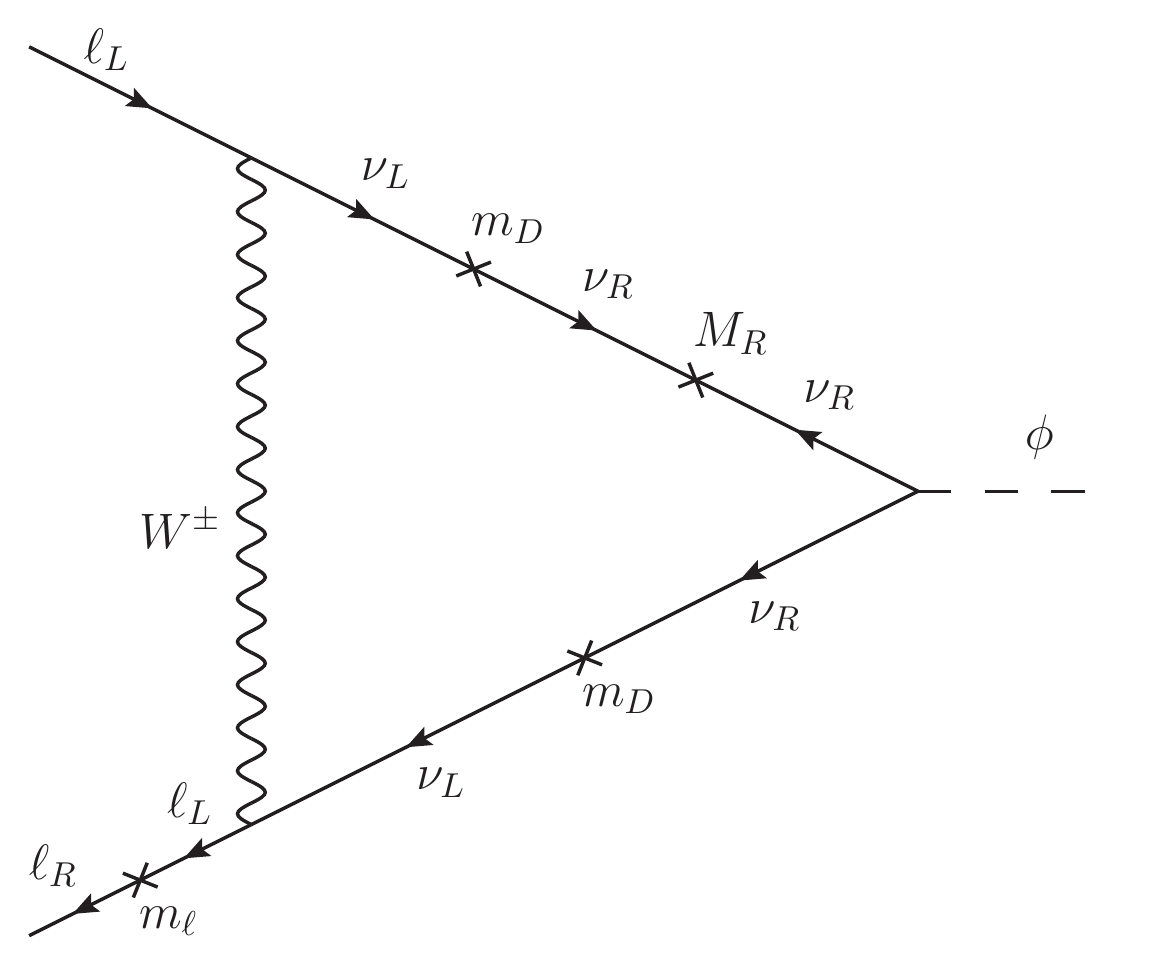}

\caption{The $W^{\pm}$-mediated loop diagram in the chiral basis, which is
equivalent to the left diagram in Fig.~\ref{fig:WZ} in the mass
basis. It shows explicitly how chirality changes in the process. Since
in the chiral basis $W^{\pm}$ only couples to left-handed leptons
and $\phi$ only  to $\nu_{R}$, we need two mass insertions of $m_{D}$
to connect $\nu_{L}$ and $\nu_{R}$. Other two mass insertions, $M_{R}$
and $m_{\ell}$, are also necessary  due to additional requirements---see
discussions in the text. \label{fig:W-chiral}}
\end{figure}

As shown in the previous section,  at tree level the scalar singlet
$\phi$ only couples to neutrinos, including light and heavy ones
in the mass basis. It does not interact with other fermions directly.
In this section, we show that one-loop corrections lead to effective
interactions of $\phi$ with charged leptons. 

From Eq.~(\ref{eq:m-10-2}), it is straightforward to check that at
the one-loop level, in the unitarity gauge (which means Goldstone
boson interactions can be ignored), there are only two possible diagrams
that can connect $\phi$ to charged leptons or quarks, as shown in
Fig.~\ref{fig:WZ}. The second diagram involving the $Z$ boson actually
leads to a pseudo-scalar coupling (see calculations later on). In
unpolarized matter, pseudo-scalar interactions cannot cause significant
long-range forces~\cite{Wilczek:1982rv,Moody:1984ba} because the
Yukawa potential between two fermions are spin dependent.  When taking
an average over the spins, the effect of pseudo-scalar interactions
vanishes. Therefore, we will focus our discussions on the first diagram
where the external fermion lines have to be charged leptons.

The diagrams in Fig.~\ref{fig:WZ} are in the mass basis which is
technically convenient for evaluation. Nonetheless it is illuminating
to show Fig.~\ref{fig:W-chiral}, another diagram in the chiral basis
which explicitly shows how chirality changes in the process. The physical
results should be basis independent. 

Fig.~\ref{fig:W-chiral} follows directly from Eq.~(\ref{eq:m-10}),
which suggests that $\phi$ only couples to $\nu_{R}$ while $W^{\pm}$
interacts with $\nu_{L}$. Therefore, two Dirac mass insertions ($m_{D}\nu_{L}\nu_{R}$
and $m_{D}\nu_{L}^{\dagger}\nu_{R}^{\dagger}$) are necessarily introduced
to connect $\nu_{R}$ and $\nu_{L}$, or $\nu_{R}^{\dagger}$ and
$\nu_{L}^{\dagger}$. Note that the two $W^{\pm}$ vertices have to
be conjugate to each other, which implies that from the $W^{\pm}$
side, a pair of $\nu_{L}$ and $\nu_{L}^{\dagger}$ is provided. On
the other hand, the Yukawa vertex couples $\phi$ to two $\nu_{R}$'s
rather than a pair of $\nu_{R}$ and $\nu_{R}^{\dagger}$. So a Majorana
mass insertion is required to flip the lepton number and convert one
of them to $\nu_{R}^{\dagger}$. The direction of lepton-number flow
in this diagram are represented by the arrows. Note that according
to the conventions in Sec.~\ref{sub:Notations}, $\nu_{L}$ and $\nu_{R}$
have opposite lepton numbers. So for $\nu_{R}\nu_{R}\phi$, the arrow
of $\nu_{R}$ should be outgoing. In contrast to that,  the arrow
of $\nu_{L}$ in the $W_{\mu}^{-}\ell_{L}^{\dagger}\overline{\sigma}^{\mu}\nu_{L}$
vertex goes inwardly. Finally, there should be a mass insertion of
$m_{\ell}\ell_{L}\ell_{R}$ on one of the external fermion lines because
it is impossible to write down an effective operator that consists
of $\phi$ and two $\ell_{L}$'s\,---\,the operator $\phi\ell_{L}\ell_{L}$
is not allowed due to electric charge conservation.

The chirality analysis in Fig.~\ref{fig:W-chiral} indicates that
the diagram would be proportional to $m_{D}{}^{2}M_{R}m_{\ell}$ if
all these masses are sufficiently small. If $M_{R}$ is much larger
than the typical scale of the loop momentum, then the propagators
of $\nu_{R}$ also contribute an additional factor of $M_{R}^{-2}$.
In this case, the diagram is expected to be proportional to $m_{D}{}^{2}M_{R}^{-1}m_{\ell}\sim m_{\nu}m_{\ell}$
where $m_{\nu}$ is the light neutrino mass.

Now let us compute the loop diagrams explicitly. Using the Dirac/Majorana
spinor representation in Eq.~(\ref{eq:m-20}), we can write down the
amplitudes of the two diagrams in Fig.~\ref{fig:WZ}:
\begin{eqnarray}
i{\cal M}_{W} & = & (i)^{3}\int\frac{d^{4}k}{(2\pi)^{4}}\overline{u(p_{2})}g_{W}^{j}\gamma_{L}^{\mu}\Delta_{j}(p_{j})\frac{y_{R}^{ji}P_{L}+y_{R}^{ji*}P_{R}}{2}\Delta_{i}(p_{i})g_{W}^{i*}\gamma_{L}^{\nu}u(p_{1})\Delta_{\mu\nu}^{W}(k),\label{eq:m-25}\\
i{\cal M}_{Z} & = & (i)^{3}\int\frac{d^{4}p_{i}}{(2\pi)^{4}}\overline{u(p_{2})}g_{Z}^{(\ell)}\gamma_{L}^{\mu}u(p_{1}){\rm tr}\left[-g_{Z}^{ij}\gamma_{L}^{\nu}\Delta_{j}(p_{j})\frac{y_{R}^{ji}P_{L}+y_{R}^{ji*}P_{R}}{2}\Delta_{i}(p_{i})\right]\Delta_{\mu\nu}^{Z}(q),\label{eq:m-26}
\end{eqnarray}
where $(i)^{3}$ comes from three vertices;  $p_{1}$ and $p_{2}$
are the momenta of the upper and lower external fermion lines; $p_{i}$
and $p_{j}$ are the momenta of $\nu_{i}$ and $\nu_{j}$; $q=p_{2}-p_{1}=p_{j}-p_{i}$;
$k$ is the momentum of $W$ propagator; and $g_{Z}^{(\ell)}$ is
the gauge coupling of $Z$ to the charge fermion $\ell$. The symbol
$\Delta$ denotes propagators. For Majorana spinors in the mass basis,
their propagators have the same form as Dirac propagators:
\begin{equation}
\Delta_{i}(p)=\frac{i}{\slashed{p}-m_{i}}.\label{eq:m-24}
\end{equation}
The propagators of $W^{\pm}$ and $Z$ are gauge dependent. Most generally,
in $R_{\xi}$ gauges, they are:
\begin{eqnarray}
\Delta_{\mu\nu}^{W}(k) & = & \frac{-i}{k^{2}-m_{W}^{2}}\left[g_{\mu\nu}-\frac{k_{\mu}k_{\nu}}{k^{2}-\xi m_{W}^{2}}(1-\xi)\right],\label{eq:m-22}\\
\Delta_{\mu\nu}^{Z}(k) & = & \frac{-i}{k^{2}-m_{Z}^{2}}\left[g_{\mu\nu}-\frac{k_{\mu}k_{\nu}}{k^{2}-\xi m_{Z}^{2}}(1-\xi)\right].\label{eq:m-23}
\end{eqnarray}
 The unitarity gauge corresponds to $\xi\rightarrow\infty$. Except
for the unitarity gauge, other gauges with finite $\xi$, e.g., the
Feynman-'t Hooft gauge ($\xi=1$) and the Lorentz gauge ($\xi=0$),
require the inclusion of Goldstone boson diagrams. The unitarity gauge,
albeit involving fewer diagrams by virtue of infinitely large masses
of Goldstone boson propagators, has a disadvantage in that the cancellation
of UV divergences  is less obvious---see discussions in Sec.~\ref{sub:Cancellation-of-UV}.
Nonetheless, it is straightforward to compute $i{\cal M}_{W}$ and
$i{\cal M}_{Z}$ for general values of $\xi$. 

First, let us  inspect the $i{\cal M}_{Z}$ amplitude. The loop integral
of the trace part gives rise to a quantity proportional to $q^{\nu}$:
\begin{equation}
\int\frac{d^{4}p_{i}}{(2\pi)^{4}}{\rm tr}\left[\gamma_{L}^{\nu}\Delta_{j}(p_{j})P_{L/R}\Delta_{i}(p_{i})\right]\propto q^{\nu},\label{eq:m-28}
\end{equation}
which can be expected from Lorentz invariance, explained as follows.
On the left-hand side of Eq.~(\ref{eq:m-28}) there are only two independent
momenta $p_{j}=p_{i}+q$ and $p_{i}$. After $p_{i}$ being integrated
out, the only quantity that can carry a Lorentz index is $q$ so the
result is proportional to $q^{\nu}$. Now plugging this in Eq.~(\ref{eq:m-26}),
we can immediately get a $\gamma^{5}$ sandwiched between $\overline{u(p_{2})}$
and $u(p_{1})$:
\begin{eqnarray}
\overline{u(p_{2})}\slashed{q}P_{L}u(p_{1}) & = & \overline{u(p_{2})}(\slashed{p}_{2}P_{L}-P_{R}\slashed{p}_{1})u(p_{1})\nonumber \\
 & = & m_{\ell}\overline{u(p_{2})}(P_{L}-P_{R})u(p_{1})\nonumber \\
 & = & -m_{\ell}\overline{u(p_{2})}\gamma^{5}u(p_{1}).\label{eq:m-37}
\end{eqnarray}
Therefore, the $Z$-mediated diagram induces a pseudo-scalar coupling,
which is computed in Appendix~\ref{sec:gamma5}. 

The $i{\cal M}_{W}$ amplitude can be computed by splitting the $W^{\pm}$
propagator in Eq.~(\ref{eq:m-22}) to two parts:
\begin{equation}
\Delta_{\mu\nu}^{W}(k)=-i\frac{g_{\mu\nu}-k_{\mu}k_{\nu}/m_{W}^{2}}{k^{2}-m_{W}^{2}}-i\frac{k_{\mu}k_{\nu}/m_{W}^{2}}{k^{2}-\xi m_{W}^{2}},\label{eq:m-27}
\end{equation}
where the first part does not contain $\xi$ and the second part is
important for cancellation of UV divergences. Note that when computing
Eq.~(\ref{eq:m-25}), because of the chiral projectors in $y_{R}^{ji}P_{L}+y_{R}^{ji*}P_{R}$,
the product of Dirac matrices gives
\begin{equation}
\gamma_{L}^{\mu}\frac{\slashed{p}_{j}+m_{j}}{p_{j}^{2}-m_{j}^{2}}\left[y_{R}^{ji}P_{L}+y_{R}^{ji*}P_{R}\right]\thinspace\frac{\slashed{p}_{i}+m_{i}}{p_{i}^{2}-m_{i}^{2}}\gamma_{L}^{\nu}=\gamma_{L}^{\mu}\frac{\slashed{p}_{j}m_{i}y_{R}^{ji*}+y_{R}^{ji}m_{j}\slashed{p}_{i}}{(p_{j}^{2}-m_{j}^{2})(p_{i}^{2}-m_{i}^{2})}\gamma_{L}^{\nu}.\label{eq:m-36}
\end{equation}
It implies that if $m_{i}\rightarrow0$ and $m_{j}\rightarrow0$,
the result would be zero, which agrees with our analysis in the chiral
basis.

With the above details being noted, we compute\footnote{We use {\tt Package-X} \cite{Patel:2015tea} to compute  loop integrals
analytically and our code is available from [\url{https://github.com/xunjiexu/vR_loop}].}  Eq.~(\ref{eq:m-25}) in the soft scattering limit ($q\rightarrow0$)
with the approximation of $m_{\ell}\ll m_{W}$ and obtain:
\begin{equation}
i{\cal M}_{W}=i\frac{m_{\ell}G^{ij}}{256\pi^{2}m_{W}^{2}}\left[F_{1}(m_{i},\ m_{j})+F_{2}(m_{i},\ m_{j})\right]\overline{u(p_{2})}u(p_{1})+i\lambda_{\phi\ell\ell}^{(W)}\overline{u(p_{2})}i\gamma^{5}u(p_{1}),\label{eq:m-29}
\end{equation}
where
\begin{equation}
G^{ij}\equiv g_{W}^{i*}g_{W}^{j}(m_{j}y_{R}^{ij}+m_{i}y_{R}^{ij*})=\frac{g^{2}c_{\theta}^{2}s_{\theta}^{2}}{2}\left[\begin{array}{cc}
-m_{1}(y_{R}+y_{R}^{*}) & m_{1}y_{R}^{*}-m_{4}y_{R}\\
m_{1}y_{R}-m_{4}y_{R}^{*} & m_{4}(y_{R}+y_{R}^{*})
\end{array}\right],\label{eq:m-48}
\end{equation}
and $F_{1}$ and $F_{2}$ correspond to the contributions of the first
and second parts of the $W^{\pm}$ propagator in Eq.~(\ref{eq:m-27}),
respectively. Their explicit forms are given in Appendix~\ref{sec:FF}.
The second term of Eq.~(\ref{eq:m-29}) leads to pseudo-scalar couplings
which cannot cause significant effect in unpolarized matter. Nevertheless,
we compute the loop-induced pseudo-scalar couplings in Appendix~\ref{sec:gamma5}.

We need to sum over $i$ and $j$ in Eq.~(\ref{eq:m-29}) to get a
finite and gauge independent result. There are several cancellations
involved in the summation, which are discussed in detail in Appendix~\ref{sec:Some-Cancellations}.
After a careful treatment of these cancellations, we obtain:
\begin{equation}
i{\cal M}_{W}\approx i\overline{u(p_{2})}y_{\phi\ell\ell}u(p_{1}),\label{eq:m-54}
\end{equation}
with
\begin{equation}
y_{\phi\ell\ell}=-\frac{3G_{F}m_{1}m_{\ell}{\rm Re}(y_{R})}{16\sqrt{2}\pi^{2}}.\label{eq:m-55}
\end{equation}
It implies that the loop diagram generates the effective interaction
\begin{equation}
{\cal L}\supset y_{\phi\ell\ell}\phi\overline{\psi_{\ell}}\psi_{\ell},\label{eq:m-56}
\end{equation}
where the effective coupling $y_{\phi\ell\ell}$, given in Eq.~(\ref{eq:m-55}),
 is suppressed by the neutrino mass $m_{\nu}$ and the charged lepton
mass $m_{\ell}$.

\section{Generalization to three flavors\label{sec:3nu-1}}

So far we have only considered leptons of a single flavor for which
we have computed the loop-induced coupling $y_{\phi\ell\ell}$, as
given in Eq.~(\ref{eq:m-55}). Now we would like to generalize it
to the realistic scenario with three flavors.

Assuming there are three generations of $\nu_{L}$ and $\nu_{R}$,
we can express the neutrino mass terms in a similar way to Eq.~(\ref{eq:m-4})
except that now the mass matrix is interpreted as a  $6\times6$
matrix:
\begin{equation}
M_{6\nu}=\left[\begin{array}{cc}
0 & m_{D}\\
m_{D}^{T} & M_{R}
\end{array}\right]_{6\times6},\label{eq:m-57-1}
\end{equation}
where $m_{D}$ and $M_{R}$ are $3\times3$ Dirac and Majorana mass
matrices respectively. In principle, the number of right-handed neutrinos
does not have to be three. It can be two or more. But to make it concrete,
let us concentrate on the case with three $\nu_{L}$ plus three $\nu_{R}$.

The neutrino mass terms and Yukawa terms are formulated as:
\begin{equation}
{\cal L}\supset\frac{1}{2}(\nu_{L}^{T},\ \nu_{R}^{T})M_{6\nu}\left(\begin{array}{c}
\nu_{L}\\
\nu_{R}
\end{array}\right)+\frac{1}{2}\phi\nu_{R}^{T}Y_{R}^{0}\nu_{R}+{\rm h.c.},\label{eq:m-84-1}
\end{equation}
where $Y_{R}^{0}$ is a $3\times3$ Yukawa coupling matrix.

A detailed analysis of this scenario is delegated to Appendix~\ref{sec:3nu-app}.
Here we simply summarize the results. In general, without any requirements
of $m_{D}$, $M_{R}$ and $Y_{R}^{0}$, the loop-induced coupling
can be numerically obtained from
\begin{equation}
y_{\phi\ell\ell}=\frac{G_{F}m_{\ell}}{64\sqrt{2}\pi^{2}}\sum_{i,\thinspace j}U_{\ell i}^{*}U_{\ell j}\left(Y_{R}M_{d}+M_{d}Y_{R}^{\dagger}\right)_{ij}F_{12}(m_{i},\ m_{j}),\label{eq:m-63}
\end{equation}
where $F_{12}$ can be computed using Eq.~(\ref{eq:m-43}), $U$ is
the full $6\times6$ mixing matrix that can diagonalize $M_{6\nu}$,
$M_{d}\equiv U^{T}M_{6\nu}U={\rm diag}(m_{1},\ m_{2},\ \cdots,\ m_{6})$
is the diagonalized form of $M_{6\nu}$, and $Y_{R}$ is the mass-basis
form of $Y_{R}^{0}$:
\begin{equation}
Y_{R}\equiv U^{T}{\rm diag}(0_{3\times3},\ Y_{R}^{0})U.\label{eq:m-91}
\end{equation}

If $M_{R}$ and $Y_{R}^{0}$ can be simultaneously editorialized\footnote{Such a feature could arise from flavor symmetries, see models in Refs.~\cite{Smirnov:2018luj,Rodejohann:2017lre,Rodejohann:2015hka}.},
then without loss of generality, we can assume $M_{R}$ and $Y_{R}^{0}$
are diagonal. Under this assumption, the result can be further simplified
to
\begin{equation}
y_{\phi\ell\ell}\approx-\frac{3G_{F}m_{\ell}}{32\sqrt{2}\pi^{2}}\left[m_{D}(Y_{R}^{0}+Y_{R}^{0\dagger})M_{R}^{-1}m_{D}^{\dagger}\right]_{\ell\ell}.\label{eq:m-90}
\end{equation}

Eq.~(\ref{eq:m-90}) can also be expressed in the Casas-Ibarra parametrization~\cite{Casas:2001sr}:
\begin{equation}
y_{\phi\ell\ell}\approx-\frac{3G_{F}m_{\ell}}{32\sqrt{2}\pi^{2}}\left[U_{L}^{*}\sqrt{m_{\nu}^{d}}R^{T}(Y_{R}^{0}+Y_{R}^{0\dagger})R^{*}\sqrt{m_{\nu}^{d}}U_{L}^{T}\right]_{\ell\ell},\label{eq:m-92}
\end{equation}
where $U_{L}$ is the PMNS matrix, $m_{\nu}^{d}={\rm diag}(m_{1},\ m_{2},\ m_{3})$,
and $R$ is a complex orthogonal matrix ($RR^{T}=1$), which is determined
by $m_{D}=iU_{L}^{*}\sqrt{m_{\nu}^{d}}R^{T}\sqrt{M_{R}^{-1}}$ in
the Casas-Ibarra parametrization. Note that our convention of $U_{L}$
is chosen in the  way that $U_{L}^{T}m_{\nu}U_{L}=m_{\nu}^{d}$ for
$m_{\nu}\equiv-m_{D}M_{R}^{-1}m_{D}^{T}$.

\section{Phenomenology\label{sec:Phenomenology}}

The loop-induced interaction of $\phi$ with electrons leads to a
Yukawa potential between two objects containing $N_{1}$ and $N_{2}$
electrons,
\begin{equation}
V(r)=-\frac{y_{\phi ee}^{2}N_{1}N_{2}}{4\pi r}e^{-m_{\phi}r}.\label{eq:m-76}
\end{equation}
 The effective Yukawa coupling $y_{\phi ee}$ is of order $G_{F}m_{e}m_{\nu}/(16\pi^{2})\sim{\cal O}(10^{-21})$,
which reaches the current sensitivity of long-range force searches.
If we replace electrons with muons, the effective coupling is generally
two orders of magnitude larger because $m_{\mu}/m_{e}\approx200$.
The muonic long-range force can be tested in binary systems of neutron
stars (NS) which contain ${\cal O}(0.1\sim1\%)$ muons of the total
mass~\cite{Pearson:2018tkr}. In particular, the recent gravitational
wave observation of NS binary mergers by the LIGO collaboration~\cite{TheLIGOScientific:2017qsa,Abbott:2020khf}
are able to test the  muonic force with unprecedented sensitivity.

As indicated by Eq.~(\ref{eq:m-63}), the value of $y_{\phi\ell\ell}$
depends on neutrino masses and the Yukawa couplings of $\phi$ to
$\nu_{R}$.  Since there are many free parameters in  $Y_{R}$  and
$M_{\nu}$ (where Majorana phases, the Dirac CP phase, the lightest
neutrino mass are still unknown), we would like to simply parametrize
$y_{\phi\ell\ell}$  as follows:
\begin{eqnarray}
y_{\phi ee} & = & \frac{3G_{F}m_{e}Y_{R}^{(e)}m_{\nu}^{(e)}}{16\sqrt{2}\pi^{2}}\approx8.0\times10^{-22}\thinspace Y_{R}^{(e)}\left(\frac{m_{\nu}^{(e)}}{0.01\ {\rm eV}}\right),\label{eq:m-81}\\
y_{\phi\mu\mu} & = & \frac{3G_{F}m_{\mu}Y_{R}^{(\mu)}m_{\nu}^{(\mu)}}{16\sqrt{2}\pi^{2}}\approx5.0\times10^{-19}\thinspace Y_{R}^{(\mu)}\left(\frac{m_{\nu}^{(\mu)}}{0.03\ {\rm eV}}\right),\label{eq:m-80}
\end{eqnarray}
where $Y_{R}^{(e)}$ and $Y_{R}^{(\mu)}$ account for the suppression
caused by the original Yukawa couplings if they are not of ${\cal O}(1)$,
while $m_{\nu}^{(e)}$ and $m_{\nu}^{(\mu)}$ account for the suppression
due to neutrino masses. In the limit of $Y_{R1}=Y_{R2}=Y_{R3}$ and
$U_{{\rm PMNS}}^{*}=U_{{\rm PMNS}}$, $m_{\nu}^{(e)}$ would be identical
to the neutrino mass matrix element responsible for neutrinoless double
beta decay (often denoted as $m_{ee}$ in the literature). But in
general,  they are different. Since $Y_{R}^{(e)}m_{\nu}^{(e)}$ and
$Y_{R}^{(\mu)}m_{\nu}^{(\mu)}$ depend on a lot of unknown fundamental
parameters, it is possible that the Majorana phases and other free
parameters conspire in such a way that $Y_{R}^{(e)}m_{\nu}^{(e)}=0$
while $Y_{R}^{(\mu)}m_{\nu}^{(\mu)}$ is not suppressed or vice versa,
analogous to the well-known fact that $m_{ee}$ for neutrinoless double
beta decay can vanish in the normal mass ordering.

\begin{figure}[t]
\centering

\includegraphics[width=0.8\textwidth]{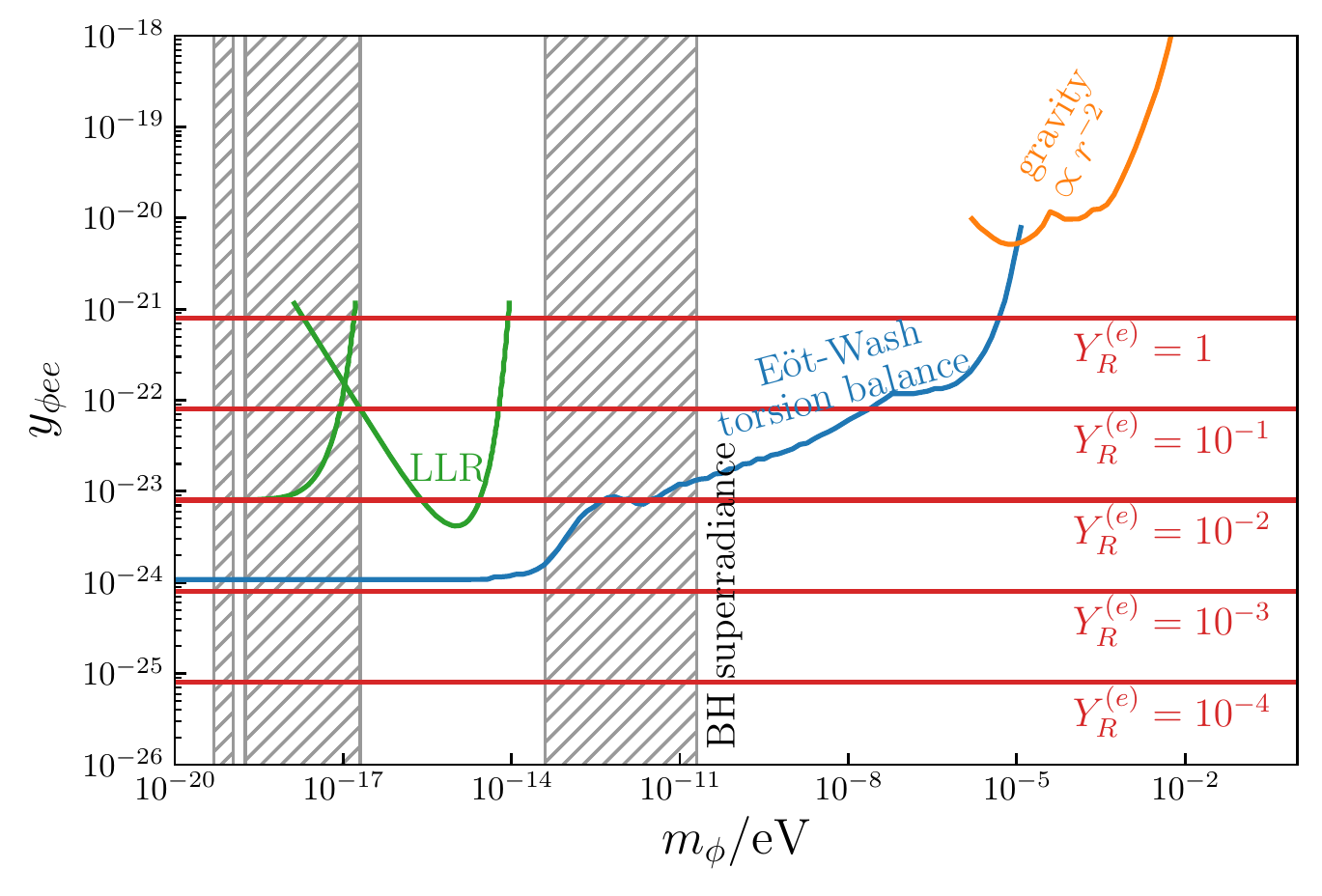}

\caption{The effective Yukawa coupling of $\phi$ to $e$,  compared with
experimental limits. The predictions of our model (red) are evaluated
according to Eq.~(\ref{eq:m-81}) with $m_{\nu}^{(e)}=0.01$ eV. The
experimental limits come from the E\"ot-Wash torsion-balance tests
of the equivalence principle (blue)~\cite{Wagner:2012ui}, tests
of gravitational inverse-square law (orange)~\cite{Adelberger:2009zz},
lunar laser-ranging (LLR, green) measurements~\cite{Wagner:2012ui,Turyshev:2006gm},
and black hole superradiance (hatched bands)~\cite{Baryakhtar:2017ngi}.\label{fig:Yee}}
\end{figure}

\begin{figure}[t]
\centering

\includegraphics[width=0.8\textwidth]{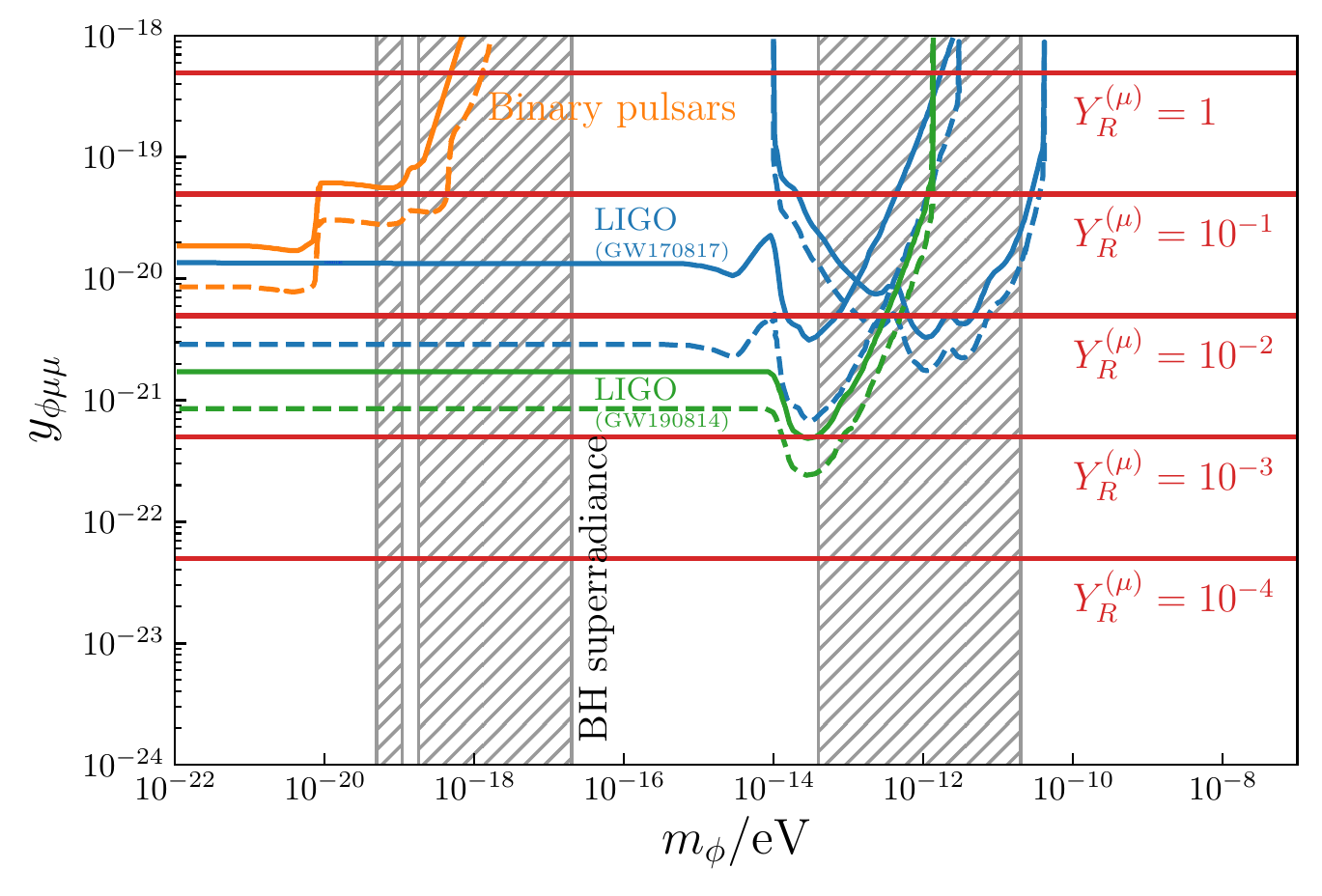}

\caption{The effective Yukawa coupling of $\phi$ to $\mu$,  compared with
experimental limits. The predictions of our model (red) are evaluated
according to Eq.~(\ref{eq:m-80}) with $m_{\nu}^{(\mu)}=0.03$ eV.
The muonic force could be probed in binary systems of neutron stars
(NS) due to the considerable abundance of muons.  The blue and green
curves represent current sensitivity of the LIGO observations of GW170817
(NS-NS merger) and GW190814 (NS-BH merger) events, respectively. Solid
(dashed) curves take conservative (optimistic) estimates of the muon
abundance~\cite{Dror:2019uea}. In addition, precision measurements
of binary pulsar systems are also sensitive to the muonic force (orange
curves)~\cite{Dror:2019uea}. \label{fig:Ymumu}}
\end{figure}

Next, we shall confront the theoretical predictions with experimental
limits, as shown in Figs.~\ref{fig:Yee} and \ref{fig:Ymumu} for
$y_{\phi ee}$ and $y_{\phi\mu\mu}$ respectively.

For $y_{\phi ee}$, current limits come from long-range force searches
of normal matter, which  have long been investigated in precision
tests of gravity,  in particular, in tests of the equivalence principle.
The Yukawa force mediated by $\phi$  can affect the former by 
contributing an exponential term to the total force and affect the
latter due to its leptophilic coupling, which causes differential
free-fall accelerations for different materials. So far, the E\"ot-Wash
torsion-balance experiment has performed tests of the weak equivalence
principle with the highest precision~\cite{Schlamminger:2007ht,Heckel:2008hw},
leading to the most stringent constraint on $y_{\phi ee}$  in the
regime of very small $m_{\phi}$. In addition, the lunar laser-ranging
(LLR) technology which is able to measure the varying distance between
the moon and the earth to high precision using laser pulses is also
sensitive to new long-range forces~\cite{Turyshev:2006gm}.  These
two bounds, reviewed in Ref.~\cite{Wagner:2012ui}, are presented
in Fig.~\ref{fig:Yee} and overlap with the theoretically most favored
region (red lines). 

For larger masses, $y_{\phi ee}$ is constrained by tests of the inverse-square
law of gravity~\cite{Adelberger:2006dh,Adelberger:2009zz}, the Casimir
effect~\cite{Bordag:2001qi}, stellar cooling processes~\cite{Davidson:2000hf,Redondo:2013lna},
$N_{{\rm eff}}$ in cosmology~\cite{Boehm:2012gr, Kamada:2015era, Huang:2017egl,Kamada:2018zxi,Luo:2020sho},
supernovae~\cite{Choi:1987sd,Choi:1989hi,Kachelriess:2000qc,Hannestad:2002ff,Farzan:2002wx,Dent:2012mx,Dreiner:2013mua},
neutrino scattering~\cite{Bilmis:2015lja,Lindner:2016wff,Farzan:2018gtr,Lindner:2018kjo,Khan:2019jvr,Link:2019pbm},
etc. But all these bounds are significantly higher than the largest
expected values of $y_{\phi ee}$---see Ref.~\cite{Heeck:2014zfa}
for a recent compilation of these bounds.

In Fig.~\ref{fig:Yee} (also Fig.~\ref{fig:Ymumu}), we add  hatched
bands to represent the constraint from black hole superradiance~\cite{Baryakhtar:2017ngi},
which is  independent of the Yukawa couplings because the effect is
caused by $\phi$ coupling to the spacetime. 

For $y_{\phi\mu\mu}$, the aforementioned laboratory constraints do
not apply since normal matter does not contain muons. Neutron stars,
 however, can be a powerful probe of muonic forces due to a significant
abundance muons in them, which is expected when the Fermi energy 
exceeds the muon mass. According to the calculations in Refs.~\cite{Pearson:2018tkr,Dror:2019uea},
 the number density of muons is typically of ${\cal O}(1\sim10)\%$
of the total number density, which is lower than but still comparable
to the electron number density\footnote{See Fig.~23 in Ref.~\cite{Pearson:2018tkr} and Fig.~3 in Ref.~\cite{Dror:2019uea}.
In the former, the number densities of protons and electrons are presented.
Assuming charge neutrality of the NS, the difference between  proton
and electron number densities  is approximately the muon number density.
 The latter needs to be converted from mass ratios to number density
ratios by multiplying a factor of $m_{\mu}/m_{n}$ where $m_{n}$
is the neutron mass.}. 

In fact, since the electron and muon number densities are of the same
order of magnitude while $y_{\phi\ell\ell}\propto m_{\ell}$, for
NS binaries we have
\begin{equation}
F^{(\mu)}\sim\left(\frac{m_{\mu}}{m_{e}}\right)^{2}F^{(e)}\gg F^{(e)},\label{eq:m-82}
\end{equation}
where $F^{(\mu)}$ and $F^{(e)}$ are the forces caused by muons and
electrons respectively. 

The recent observations of NS-NS and NS-BH mergers by the LIGO collaboration
provide very promising data to probe the muonic force in this model.
For a NS-NS merger, the effect of $\phi$ is two-fold~\cite{Kopp:2018jom}.
First, the attractive force affects the orbital dynamics in a classical
way, i.e., modifying the Kepler\textquoteright s law when $r\sim m_{\phi}^{-1}$.
  Second, since $\phi$ is a ultra-light boson, there is radiation
of $\phi$ due to  the rotating dipole, which causes extra  energy
loss. For a NS-BH merger, only the effect of $\phi$ radiation is
relevant. An in-depth analysis of the sensitivity to muonic forces
 based on the recent two events GW170817 (NS-NS merger) and GW190814
(NS-BH merger) has been performed in Ref.~\cite{Dror:2019uea}. Their
results have been incorporated in Fig.~\ref{fig:Ymumu}, where solid
(dashed) curves are derived using a conservative (optimistic) estimate
of the muon abundance. For GW170817, the sensitivity curves of the
two effects are evaluated and presented separately. The  first effect
(orbital dynamics) are more sensitive than the second to $m_{\phi}$
when it is in the large-mass ($10^{-12}\sim10^{-10}$ eV) regime.
  In addition to binary mergers, precision measurements of binary
pulsars can also be sensitive to muonic forces ~\cite{Poddar:2019wvu,Dror:2019uea}.

As shown in Fig.~\ref{fig:Ymumu}, the LIGO curves cross the red
lines of  $Y_{R}^{(\mu)}=10^{-3}\sim1$, which implies that the loop-induced
muonic force in this model could be probed in the theoretically most
favored regime. Future experiments such as the Einstein Telescope\footnote{See the ET conceptual design document:~\url{https://tds.virgo-gw.eu/?call_file=ET-0106C-10.pdf}.}
and Cosmic Explorer~\cite{Reitze:2019iox} can substantially improve
the sensitivity to muonic forces and thus have great potential of
probing this scenario.

\section{Conclusions and Discussions\label{sec:Conclusion}}

The $\nu_{R}$-philic scalar model naturally gives rise to extremely
small couplings of charged leptons to a long-range force mediator
via loop-level processes. The small values of the loop-induced couplings
coincidentally meet the current sensitivity of long-range force searches
in laboratories and in astrophysical observations such as the recent
detection of GW from NS mergers by LIGO, as we have shown in Figs.~\ref{fig:Yee}
and \ref{fig:Ymumu}.

In this  model, loop-induced couplings to quarks also exist, due  to
the $Z$-mediated diagram in Fig.~\ref{fig:WZ}. However, our calculation
shows that only pseudo-scalar couplings are generated in this case,
the effect of which is suppressed in unpolarized matter.

Our loop calculation result for the most general three-flavor case
is given by Eq.~(\ref{eq:m-63}) which, though involving diagonalization
of the full $6\times6$ mass matrix, can be numerically evaluated.
For the special case where $M_{R}$ and $Y_{R}^{0}$ can be simultaneously
diagonalized, the result can be further simplified to Eq.~(\ref{eq:m-92}),
where the dependence on the PMNS matrix is manifestly extracted. 

Our results can also be used to obtain loop-induced interactions for
other similar models that contain the diagrams in Fig.~\ref{fig:WZ},
via proper replacements of the couplings in vertices and masses in
propagators. However, one caveat should be noted here that incomplete
models where the tree-level couplings of $\phi$ to light neutrino
mass eigenstates are not governed by the active-sterile neutrino mixing
would lead to gauge dependent results.
\begin{acknowledgments}
We thank Andreas Trautner and Toby Opferkuch for useful discussions.
\end{acknowledgments}

\appendix

\section{Full expressions of $F_{1}$ and $F_{2}$ \label{sec:FF}}

The explicit expressions of $F_{1}$ and $F_{2}$ read as follows:
\begin{eqnarray}
F_{1} & = & 6\left(\frac{1}{\epsilon}+\log\frac{\mu^{2}}{m_{W}^{2}}\right)-2\frac{m_{j}^{2}+2m_{W}^{2}}{m_{j}^{2}-m_{W}^{2}}\log\frac{m_{j}^{2}}{m_{W}^{2}}\nonumber \\
 &  & +\frac{5m_{i}^{2}m_{j}^{2}-5m_{i}^{2}m_{W}^{2}-5m_{j}^{2}m_{W}^{2}+11m_{W}^{4}}{\left(m_{i}^{2}-m_{W}^{2}\right)\left(m_{j}^{2}-m_{W}^{2}\right)}\nonumber \\
 &  & -\frac{2\left(m_{i}^{2}m_{j}^{2}m_{W}^{2}+m_{i}^{2}m_{j}^{4}-2m_{i}^{2}m_{W}^{4}-7m_{j}^{4}m_{W}^{2}+2m_{j}^{2}m_{W}^{4}+2m_{j}^{6}\right)\log\frac{m_{i}^{2}}{m_{j}^{2}}}{\left(m_{i}^{2}-m_{j}^{2}\right)\left(m_{j}^{2}-m_{W}^{2}\right){}^{2}}\nonumber \\
 &  & -\frac{2m_{W}^{4}\left(17m_{i}^{2}m_{j}^{2}-10m_{i}^{2}m_{W}^{2}+5m_{i}^{4}-7m_{j}^{2}m_{W}^{2}+2m_{j}^{4}+2m_{W}^{4}\right)\log\frac{m_{i}^{2}}{m_{W}^{2}}}{\left(m_{i}^{2}-m_{W}^{2}\right){}^{2}\left(m_{j}^{2}-m_{W}^{2}\right){}^{2}}\nonumber \\
 &  & -\frac{2m_{i}^{2}m_{j}^{2}\left(2m_{i}^{2}m_{j}^{2}-7m_{i}^{2}m_{W}^{2}-4m_{j}^{2}m_{W}^{2}\right)\log\frac{m_{i}^{2}}{m_{W}^{2}}}{\left(m_{i}^{2}-m_{W}^{2}\right){}^{2}\left(m_{j}^{2}-m_{W}^{2}\right){}^{2}}\thinspace,\label{eq:F1}\\
\nonumber \\
F_{2} & = & -6\left(\frac{1}{\epsilon}+\log\frac{\mu^{2}}{\xi m_{W}^{2}}\right)+2\frac{m_{j}^{2}}{m_{j}^{2}-\xi m_{W}^{2}}\log\frac{m_{j}^{2}}{\xi m_{W}^{2}}\nonumber \\
 &  & -\frac{5m_{i}^{2}\left(m_{j}^{2}-\xi m_{W}^{2}\right)+\xi m_{W}^{2}\left(7\xi m_{W}^{2}-5m_{j}^{2}\right)}{\left(m_{i}^{2}-\xi m_{W}^{2}\right)\left(m_{j}^{2}-\xi m_{W}^{2}\right)}\nonumber \\
 &  & +\frac{2m_{j}^{2}\left[m_{i}^{2}\left(m_{j}^{2}-\xi m_{W}^{2}\right)-3\xi m_{j}^{2}m_{W}^{2}+2m_{j}^{4}\right]\log\frac{m_{i}^{2}}{m_{j}^{2}}}{\left(m_{i}^{2}-m_{j}^{2}\right)\left(m_{j}^{2}-\xi m_{W}^{2}\right){}^{2}}\nonumber \\
 &  & -\frac{2\xi^{2}m_{j}^{2}m_{W}^{4}\left(2m_{j}^{2}-3\xi m_{W}^{2}\right)\log\frac{m_{i}^{2}}{\xi m_{W}^{2}}}{\left(m_{i}^{2}-\xi m_{W}^{2}\right){}^{2}\left(m_{j}^{2}-\xi m_{W}^{2}\right){}^{2}}\nonumber \\
 &  & +\frac{2\xi m_{i}^{2}m_{W}^{2}\left(9\xi m_{j}^{2}m_{W}^{2}-4m_{j}^{4}-4\xi^{2}m_{W}^{4}\right)\log\frac{m_{i}^{2}}{\xi m_{W}^{2}}}{\left(m_{i}^{2}-\xi m_{W}^{2}\right){}^{2}\left(m_{j}^{2}-\xi m_{W}^{2}\right){}^{2}}\nonumber \\
 &  & +\frac{2m_{i}^{4}\left(2m_{j}^{4}-5\xi m_{j}^{2}m_{W}^{2}+3\xi^{2}m_{W}^{4}\right)\log\frac{m_{i}^{2}}{\xi m_{W}^{2}}}{\left(m_{i}^{2}-\xi m_{W}^{2}\right){}^{2}\left(m_{j}^{2}-\xi m_{W}^{2}\right){}^{2}}\thinspace.\label{eq:F2}
\end{eqnarray}
Here we have used dimensional regularization which means the integrals
are computed in a $d=4-2\epsilon$ dimensional spacetime. And the
generalization of integration measure $\int\frac{d^{4}k}{(2\pi)^{4}}\rightarrow\mu^{2\epsilon}\int\frac{d^{d}k}{(2\pi)^{d}}$
introduces a dimensional constant $\mu$ which, together with $1/\epsilon$,
should be canceled out in physical results.

We have verified that  the above expressions are symmetric under
$i\leftrightarrow j$:
\begin{equation}
F_{1}=\left.F_{1}\right|_{i\leftrightarrow j},\ \ F_{2}=\left.F_{2}\right|_{i\leftrightarrow j}.\label{eq:m-30}
\end{equation}
In addition, though $m_{i}^{2}-m_{j}^{2}$ appears in some of the
denominators, it does not cause additional divergences when $m_{i}\rightarrow m_{j}$:
\begin{eqnarray}
\lim_{m_{j}\rightarrow m_{i}}F_{1} & = & 6\left(\frac{1}{\epsilon}+\log\frac{\mu^{2}}{m_{W}^{2}}\right)\nonumber \\
 &  & +\frac{3m_{i}^{2}m_{W}^{4}\left[3-8\log\frac{m_{i}^{2}}{m_{W}^{2}}\right]+\left(3m_{i}^{4}m_{W}^{2}-m_{i}^{6}\right)\left[1+6\log\frac{m_{i}^{2}}{m_{W}^{2}}\right]-11m_{W}^{6}}{\left(m_{i}^{2}-m_{W}^{2}\right){}^{3}},\label{eq:m-31}\\
\nonumber \\
\lim_{m_{j}\rightarrow m_{i}}F_{2} & = & -6\left(\frac{1}{\epsilon}+\log\frac{\mu^{2}}{\xi m_{W}^{2}}\right)\nonumber \\
 &  & +\frac{2m_{i}^{2}\left(-9\xi m_{i}^{2}m_{W}^{2}+3m_{i}^{4}+8\xi^{2}m_{W}^{4}\right)}{\left(m_{i}^{2}-\xi m_{W}^{2}\right){}^{3}}\log\frac{m_{i}^{2}}{\xi m_{W}^{2}}\nonumber \\
 &  & +\frac{-9\xi^{2}m_{i}^{2}m_{W}^{4}+\xi m_{i}^{4}m_{W}^{2}+m_{i}^{6}+7\xi^{3}m_{W}^{6}}{\left(m_{i}^{2}-\xi m_{W}^{2}\right){}^{3}}.\label{eq:m-32}
\end{eqnarray}
To obtain the final result of $i{\cal M}_{W}$, one needs both Eqs.~(\ref{eq:F1})-(\ref{eq:F2})
and Eqs.~(\ref{eq:m-31})-(\ref{eq:m-32}) to sum over $i$ and $j$
as it involves cases of $i\neq j$ and $i=j$.

\section{Some Cancellations\label{sec:Some-Cancellations}}

In this appendix, we discuss several noteworthy cancellations in our
calculation.

\subsection{Cancellation of UV divergences\label{sub:Cancellation-of-UV}}

As can be seen from Eqs.~(\ref{eq:F1}) and (\ref{eq:F2}), both $F_{1}$
and $F_{2}$ contain UV divergences $1/\epsilon$ in their first terms.
When combined together in Eq.~(\ref{eq:m-29}), there is obviously
a cancellation between the two divergences:
\begin{equation}
6\left(\frac{1}{\epsilon}+\log\frac{\mu^{2}}{m_{W}^{2}}\right)-6\left(\frac{1}{\epsilon}+\log\frac{\mu^{2}}{\xi m_{W}^{2}}\right)=6\log\xi.\label{eq:m-33}
\end{equation}
Further cancellations of $\log\xi$ will be discussed in the next
subsection. 

Here we would like to address a subtlety concerning  UV divergences
in the unitarity gauge. If we had naively taken the $\xi\rightarrow\infty$
limit of Eq.~(\ref{eq:m-22}) at the beginning of the above calculations,
we would get a divergent result because
\begin{equation}
\lim_{\xi\rightarrow\infty}\Delta_{\mu\nu}^{W}(k)=-i\frac{g_{\mu\nu}-k_{\mu}k_{\nu}/m_{W}^{2}}{k^{2}-m_{W}^{2}},\label{eq:m-34}
\end{equation}
which corresponds to exactly the $F_{1}$ contribution according to
Eq.~(\ref{eq:m-27}). And our calculation has shown that the $F_{1}$
contribution itself is UV divergent. We also know that the divergence
is actually canceled out by the $F_{2}$ contribution, which however
would vanish if $\xi\rightarrow\infty$ had been taken in the naive
way. That implies that taking $\xi\rightarrow\infty$ should be after
the loop integration. Actually from the second term of ~(\ref{eq:m-27}),
one can see that when the loop integral contains $\frac{k_{\mu}k_{\nu}/m_{W}^{2}}{k^{2}-\xi m_{W}^{2}}$,
the $\xi\rightarrow\infty$ limit does not commute with $k\rightarrow\infty$
in the integral. Taking $\xi\rightarrow\infty$ after the integration
can make the large momentum contribution with $k^{2}>\xi m_{W}^{2}$
be included, which is crucial for the UV cancellation. 

In other gauges, it is  more straightforward to see that $i{\cal M}_{W}$
is finite. Taking the Feynman-'t Hooft gauge for example, 
\begin{equation}
\lim_{\xi\rightarrow1}\Delta_{\mu\nu}^{W}(k)=\frac{-ig_{\mu\nu}}{k^{2}-m_{W}^{2}},\label{eq:m-35}
\end{equation}
when it is applied to Eq.~(\ref{eq:m-25}), using Eq.~(\ref{eq:m-36}),
the loop integral becomes
\begin{equation}
\int\frac{d^{4}k}{(2\pi)^{4}}\gamma_{L}^{\mu}\frac{\slashed{p}_{j}\lambda_{j}+\slashed{p}_{i}\lambda_{i}}{(p_{j}^{2}-m_{j}^{2})(p_{i}^{2}-m_{i}^{2})}\gamma_{L}^{\nu}\frac{g_{\mu\nu}}{k^{2}-m_{W}^{2}}\ \xrightarrow{{\rm \ large}\ k}\ (\lambda_{i}+\lambda_{j})\int\frac{d^{4}k}{(2\pi)^{4}}\frac{\gamma_{L}^{\mu}\slashed{k}\gamma_{L\mu}}{k^{6}},\label{eq:m-40}
\end{equation}
where $\lambda_{j}\equiv m_{i}y_{R}^{ji*}$ and $\lambda_{i}\equiv y_{R}^{ji}m_{j}$.
Is is now obvious to see that the loop integral  converges because
the integrand is proportional to $k^{-5}$ as $k\rightarrow\infty$.

\subsection{Cancellation of $\xi$ dependence in  $R_{\xi}$ gauges\label{sub:gauge-dependence}}

\begin{figure}[t]
\centering

\includegraphics[width=0.3\textwidth]{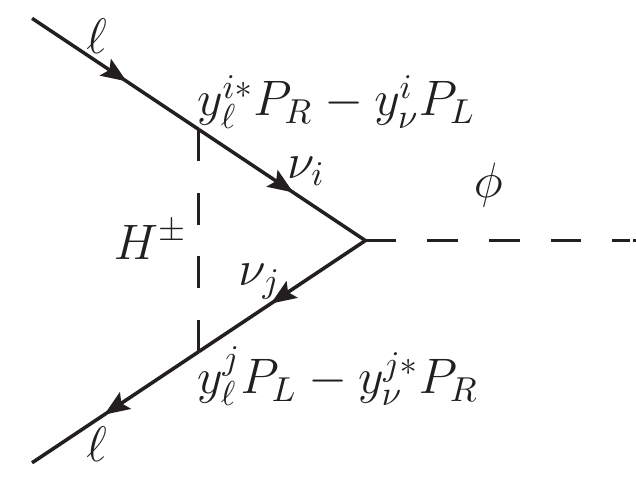}

\caption{The Goldstone boson diagram that complements the $W^{\pm}$ diagram
to cancel the $\xi$ dependence in $R_{\xi}$ gauges.  \label{fig:H-diagram}}
\end{figure}

The $F_{1}$ contribution is $\xi$ independent. So we are only concerned
with $F_{2}$. Let us make a series expansion of $F_{2}$ with respect
to $\xi^{-1}$:
\begin{equation}
F_{2}=-6\left(\frac{1}{\epsilon}+\log\frac{\mu^{2}}{m_{W}^{2}}\right)-7+6\log\xi+{\cal O}(\xi^{-1}).\label{eq:m-38}
\end{equation}
Usually in $R_{\xi}$ gauges, it is expected that the $\xi$ dependence
of a $W^{\pm}$ diagram is canceled by a similar diagram with $W^{\pm}$
replaced by its Goldstone boson $H^{\pm}$. In our case, it would
be the diagram in Fig.~\ref{fig:H-diagram}. However, a straightforward
calculation shows that the amplitude of this diagram is
\begin{equation}
i{\cal M}_{H^{\pm}}\propto m_{\ell}\frac{m_{j}y_{R}^{ij}+m_{i}y_{R}^{ij*}}{m_{W}^{2}\xi}+{\cal O}(\xi^{-2}),\label{eq:m-39}
\end{equation}
which is impossible to cancel the $\log\xi$ term in Eq.~(\ref{eq:m-38})
when $\xi$ increases to sufficiently large values.

This problem is essentially related to the completeness of the model.
For an arbitrary matrix of $y_{R}^{ij}$, indeed the result would
be gauge dependent and the $\log\xi$ term remains for each case of
$(i,j)=(1,1)$, $(1,4)$, $(4,1)$, and $(4,4)$. However,   in
Sec.~\ref{sec:basic} we have shown that the elements in $y_{R}^{ij}$
are correlated by active-sterile neutrino mixing---see Eq.~(\ref{eq:m-11}).
Besides, $g_{W}^{i}$ also depends on the mixing---see Eq.~(\ref{eq:m-12}).
 As a consequence, when summing up the contributions of both light
and heavy neutrinos in Eq.~(\ref{eq:m-29}), the $\log\xi$ term cancels
out because 
\begin{equation}
\sum_{i,\thinspace j}G^{ij}6\log\xi=0.\label{eq:m-47}
\end{equation}
Therefore,  the $\log\xi$ term can be safely ignored when computing
the full amplitude. Actually if we inspect $G^{ij}$ in the chiral
basis, the cancellation is more manifest. From Eq.~(\ref{eq:m-48}),
we can express $\sum_{i,\thinspace j}G^{ij}$ in the matrix form:
\begin{equation}
\sum_{i,\thinspace j}G^{ij}=w^{\dagger}Y_{R}\left[\begin{array}{cc}
m_{1}\\
 & m_{4}
\end{array}\right]^{\dagger}w+w^{\dagger}\left[\begin{array}{cc}
m_{1}\\
 & m_{4}
\end{array}\right]Y_{R}^{*}w,\label{eq:m-49}
\end{equation}
where $w$ and $Y_{R}$ are the vector and matrix of $g_{W}^{i}$
and $y_{R}^{ij}$ in Eqs.~(\ref{eq:m-12}) and (\ref{eq:m-11})  respectively.
They are transformed from the chiral basis by:
\begin{equation}
w\equiv\frac{g}{\sqrt{2}}U^{T}\left[\begin{array}{c}
1\\
0
\end{array}\right],\ \ Y_{R}\equiv y_{R}U^{T}\left[\begin{array}{cc}
0\\
 & 1
\end{array}\right]U.\label{eq:m-50}
\end{equation}
Therefore, in the chiral basis, we have
\begin{eqnarray}
\sum_{i,\thinspace j}G^{ij} & \propto & (1,\ 0)U^{*}\left\{ U^{T}\left[\begin{array}{cc}
0\\
 & 1
\end{array}\right]UU^{\dagger}M_{2\nu}^{*}U^{*}+U^{T}M_{\nu}UU^{\dagger}\left[\begin{array}{cc}
0\\
 & 1
\end{array}\right]U^{*}\right\} U^{T}\left[\begin{array}{c}
1\\
0
\end{array}\right]\nonumber \\
 & = & (1,\ 0)\left\{ \left[\begin{array}{cc}
0\\
 & 1
\end{array}\right]M_{2\nu}^{*}+M_{2\nu}\left[\begin{array}{cc}
0\\
 & 1
\end{array}\right]\right\} \left[\begin{array}{c}
1\\
0
\end{array}\right]\nonumber \\
 & = & 0.\label{eq:m-85}
\end{eqnarray}
Here $M_{2\nu}$ is the neutrino mass matrix in Eq.~(\ref{eq:m-4}).
It shows that the vanishing product of $w$ and $Y_{R}$ (more specifically,
$w^{\dagger}Y_{R}=0$ and $Y_{R}^{\dagger}w=0$), which is due to
the absence of $W^{\pm}$-$\nu_{R}$  and $\phi$-$\nu_{L}$ couplings,
leads to  $\sum_{i,\thinspace j}G^{ij}=0$.

\subsection{GIM-like cancellation}

If all the neutrino masses (including heavy ones) are much smaller
than $m_{W}$, when summing over $i$ and $j$ in Eq.~(\ref{eq:m-29}),
the leading-order contribution vanishes in a way similar to the
Glashow-Iliopoulos-Maiani (GIM) mechanism~\cite{Glashow:1970gm}.
At the next-to-leading order (NLO), a nonzero result can be obtained.
But in case of zero mass splitting of neutrinos, the NLO contribution
would vanish again. This is also  similar to the GIM cancellation,
where if $u$ and $c$ quarks were of equal mass, the $K^{0}\rightarrow\overline{K^{0}}$
amplitude would be zero.

Let us compute $i{\cal M}_{W}$ in the unitarity gauge. The preceding
discussion in Sec.~\ref{sub:gauge-dependence} concludes that the
$\log\xi$ term can be safely ignored in this complete model. Hence
we define
\begin{equation}
F_{12}\equiv\lim_{\xi\rightarrow\infty}\left(F_{1}+F_{2}-6\log\xi\right),\label{eq:m-43}
\end{equation}
which is finite. Then $i{\cal M}_{W}$ in the unitarity gauge can
be computed  by:
\begin{equation}
i{\cal M}_{W}=i\overline{u(p_{2})}u(p_{1})\sum_{i,\thinspace j}\frac{m_{\ell}G^{ij}F_{12}(m_{i},\ m_{j})}{256\pi^{2}m_{W}^{2}}.\label{eq:m-51}
\end{equation}
Now if we assume $m_{W}\gg m_{i}$ and $m_{j}$, $F_{12}$ can be
expanded as follows:
\begin{equation}
F_{12}=4+\frac{6m_{i}^{4}\left(1+4\log\frac{m_{i}}{m_{W}}\right)-6m_{j}^{4}\left(1+4\log\frac{m_{j}}{m_{W}}\right)}{m_{W}^{2}\left(m_{i}^{2}-m_{j}^{2}\right)}+{\cal O}\left(\frac{m_{i,j}^{4}}{m_{W}^{4}}\right).\label{eq:m-44}
\end{equation}
The constant term in $F_{12}$ does not contribute to Eq.~(\ref{eq:m-51})
due to $\sum_{i,\thinspace j}G^{ij}=0$. So the leading-order contribution
to $i{\cal M}_{W}$ vanishes. Only the second or higher-order terms
in Eq.~(\ref{eq:m-44}) further suppressed by $m_{i,j}^{2}/m_{W}^{2}$
contribute to nonzero $i{\cal M}_{W}$.

Plugging Eq.~(\ref{eq:m-44}) into Eq.~(\ref{eq:m-51}) and using
the explicit form of $G^{ij}$ in Eq.~(\ref{eq:m-48}), we obtain
\begin{equation}
\frac{i{\cal M}_{W}}{i\overline{u(p_{2})}u(p_{1})}\approx\frac{3g^{2}m_{\ell}s_{\theta}^{2}c_{\theta}^{2}(y_{R}+y_{R}^{*})\left[4\left(2m_{1}+m_{4}\right)m_{4}^{3}\log\left(\frac{m_{4}}{m_{W}}\right)+3m_{4}^{4}+4m_{1}m_{4}^{3}-(1\leftrightarrow4)\right]}{256\pi^{2}m_{W}^{4}(m_{1}+m_{4})}.\label{eq:m-52}
\end{equation}
Note that the expression in the square bracket is antisymmetric under
the interchange of $m_{1}$ and $m_{4}$. Therefore if the mass splitting
$m_{1}-m_{4}$ is zero, the NLO contribution vanishes as well, similar
to the GIM cancellation. 

In Type I seesaw, the scale of heavy neutrino masses is often assumed
to be much higher than the electroweak scale. Hence a more likely
scenario is $m_{4}\gg m_{W}\gg m_{1}$. For such a hierarchy, there
is no GIM-like cancellation, as we shall show below.

First, we need to expand $F_{12}$ in other regimes. If the diagram
contains heavy neutrinos running in the loop, we expand it with respect
to $m_{W}$:

\begin{equation}
F_{12}\approx\frac{2m_{j}^{2}\left[1-6\log\left(\frac{m_{W}}{m_{j}}\right)\right]-2m_{i}^{2}\left[1-6\log\left(\frac{m_{W}}{m_{i}}\right)\right]}{m_{i}^{2}-m_{j}^{2}},\ \ {\rm for\ }m_{W}\ll m_{i},\ m_{j}.\label{eq:m-53}
\end{equation}
If the diagram contains one light and one heavy neutrinos, we have
\begin{equation}
F_{12}\approx-2+\left[12\log\left(\frac{m_{W}}{m_{j}}\right)-\frac{6m_{W}^{2}}{m_{j}^{2}}\right]\left(1+\frac{m_{i}^{2}}{m_{j}^{2}}\right)-\frac{6m_{i}^{2}}{m_{j}^{2}},\ \ {\rm for\ }m_{i}\ll m_{W}\ll m_{j}.\label{eq:m-45}
\end{equation}
For $m_{j}\ll m_{W}\ll m_{i}$, the result can be obtained by an interchange
of $i$ and $j$ in Eq.~(\ref{eq:m-45}).

Combining the results in Eqs.~(\ref{eq:m-44}), (\ref{eq:m-53}),
and (\ref{eq:m-45}), we sum over $i$ and $j$ in Eq.~(\ref{eq:m-51}),
which gives:
\[
\frac{i{\cal M}_{W}}{i\overline{u(p_{2})}u(p_{1})}=-3g^{2}m_{\ell}m^{3}s_{\theta}^{2}{\rm Re}(y_{R})\frac{m^{4}-4m^{2}m_{W}^{2}+m_{W}^{4}\left(3+2\log\frac{m^{2}}{m_{W}^{2}}\right)}{128\pi^{2}m_{W}^{2}\left(m^{2}-m_{W}^{2}\right){}^{3}}+{\cal O}(s_{\theta}^{4}),
\]
where $m\equiv\sqrt{m_{1}^{2}+m_{4}^{2}}$. Now taking $G_{F}=\sqrt{2}g^{2}/(8m_{W}^{2})$,
$ms_{\theta}^{2}=m_{1}$, and $m\gg m_{W}$, we obtain the result
in Eqs.~(\ref{eq:m-54}) and (\ref{eq:m-55}).

\section{Pseudo-scalar couplings\label{sec:gamma5}}

As mentioned in Sec.~\ref{sec:Loop}, the $W$ diagram in Fig.~\ref{fig:WZ}
leads to loop-induced couplings of both scalar ($\overline{\psi}\phi\psi$)
and pseudo-scalar ($\overline{\psi}i\gamma^{5}\phi\psi$) forms. The
$Z$ diagram  leads to only pseudo-scalar couplings. Although the
pseudo-scalar couplings are not relevant to the phenomenology considered
in this work, we would like to present our calculation of the pseudo-scalar
couplings in this appendix.

Let us first compute the $Z$ diagram. Starting from Eq.~(\ref{eq:m-26}),
we have:
\begin{equation}
{\rm tr}\left[-g_{Z}^{ij}\gamma_{L}^{\nu}\Delta_{j}(p_{j})\frac{y_{R}^{ji}P_{L}+y_{R}^{ji*}P_{R}}{2}\Delta_{i}(p_{i})\right]=g_{Z}^{ij}\frac{m_{j}y_{R}^{ji}p_{i}^{\nu}+m_{i}y_{R}^{ji*}p_{j}^{\nu}}{\left(p_{j}^{2}-m_{j}^{2}\right)\left(p_{i}^{2}-m_{i}^{2}\right)}.\label{eq:m-95}
\end{equation}
So Eq.~(\ref{eq:m-26}) can be written as
\begin{equation}
i{\cal M}_{Z}=(i)^{3}\int\frac{d^{4}p_{i}}{(2\pi)^{4}}\overline{u(p_{2})}g_{Z}^{(\ell)}\gamma_{L}^{\mu}u(p_{1})g_{Z}^{ij}\frac{m_{j}y_{R}^{ji}p_{i}^{\nu}+m_{i}y_{R}^{ji*}p_{j}^{\nu}}{\left(p_{j}^{2}-m_{j}^{2}\right)\left(p_{i}^{2}-m_{i}^{2}\right)}\Delta_{\mu\nu}^{Z}(q).\label{eq:m-100}
\end{equation}
In the limit of $q\rightarrow0$,
\[
\Delta_{\mu\nu}^{Z}(q)\rightarrow\frac{ig_{\mu\nu}}{m_{Z}^{2}},
\]
we replace $p_{j}$ with $p_{i}+q$ and extract the $p_{i}$-independent
part out of the loop integral:
\begin{equation}
i{\cal M}_{Z}=\frac{i}{16\pi^{2}}\overline{u(p_{2})}g_{Z}^{(\ell)}\gamma_{L\mu}u(p_{1})\frac{g_{Z}^{ij}}{m_{Z}^{2}}I_{ij}^{\mu}(q),\label{eq:m-99}
\end{equation}
where the loop integral $I^{\mu}(q)$ reads:
\begin{equation}
I_{ij}^{\mu}(q)\equiv\left[\frac{i}{16\pi^{2}}\right]^{-1}\int\frac{d^{4}p_{i}}{(2\pi)^{4}}\frac{m_{j}y_{R}^{ji}p_{i}^{\mu}+m_{i}y_{R}^{ji*}(p_{i}^{\mu}+q^{\mu})}{\left[(p_{i}^{\mu}+q^{\mu})^{2}-m_{j}^{2}\right]\left(p_{i}^{2}-m_{i}^{2}\right)}.\label{eq:m-98}
\end{equation}
Performing the loop integration, we get
\begin{eqnarray}
I_{ij}^{\mu}(q) & = & q^{\mu}\frac{-m_{j}y_{R}^{ji}+m_{i}y_{R}^{ji*}}{2}\left(\frac{1}{\epsilon}+\log\mu^{2}-\log m_{j}^{2}\right)\nonumber \\
 &  & +q^{\mu}\frac{y_{R}^{ji}m_{j}(m_{j}^{2}-3m_{i}^{2})+y_{R}^{ji*}m_{i}(m_{i}^{2}-3m_{j}^{2})}{4(m_{i}^{2}-m_{j}^{2})}\nonumber \\
 &  & +q^{\mu}\frac{-y_{R}^{ji}m_{j}m_{i}+y_{R}^{ji*}(m_{i}^{2}-2m_{j}^{2})}{2(m_{i}^{2}-m_{j}^{2})^{2}}m_{i}^{3}\log\frac{m_{i}^{2}}{m_{j}^{2}}\nonumber \\
 &  & +q^{\mu}{\cal O}(q^{2}).\label{eq:m-97}
\end{eqnarray}
Next, we sum over $i$ and $j$ and expand the result in $s_{\theta}$:
\begin{equation}
\sum_{i,\thinspace j}g_{Z}^{ij}I_{ij}^{\mu}(q)\approx q^{\mu}\frac{y_{R}-y_{R}^{*}}{4}ms_{\theta}^{2}+{\cal O}(s_{\theta}^{3}).\label{eq:m-96}
\end{equation}

The UV divergence cancels out in the summation because
\begin{equation}
\sum_{i,\thinspace j}g_{Z}^{ij}m_{j}y_{R}^{ji}=\sum_{i,\thinspace j}g_{Z}^{ij}m_{i}y_{R}^{ji*}=0,\label{eq:m-101}
\end{equation}
which can be proven straightforwardly from Eqs.~(\ref{eq:m-11}). 

Plugging Eq.~(\ref{eq:m-96}) into Eq.~(\ref{eq:m-99}), we obtain
\begin{eqnarray}
i{\cal M}_{Z} & = & i\frac{g_{Z}^{(\ell)}m_{1}(y_{R}-y_{R}^{*})}{64\pi^{2}m_{Z}^{2}}\overline{u(p_{2})}\slashed{q}P_{L}u(p_{1})\label{eq:m-102}\\
 & = & -i\frac{g_{Z}^{(\ell)}m_{\ell}m_{1}(y_{R}-y_{R}^{*})}{64\pi^{2}m_{Z}^{2}}\overline{u(p_{2})}\gamma^{5}u(p_{1}),\label{eq:m-105}
\end{eqnarray}
where in the second step we have used Eq.~(\ref{eq:m-37}).

Computing the pseudo-scalar coupling from the $W$ diagram is similar,
except that the bilinear $\overline{u(p_{2})}(C_{1}\slashed{p}_{1}+C_{2}\slashed{p}_{2})P_{L}u(p_{1})$
where $C_{1}$ and $C_{2}$ are different scalar quantities cannot
be converted to $\overline{u(p_{2})}\slashed{q}P_{L}u(p_{1})$. It
actually contributes to both scalar and pseudo-scalar couplings because
\begin{eqnarray*}
\overline{u(p_{2})}(C_{1}\slashed{p}_{1}+C_{2}\slashed{p}_{2})P_{L}u(p_{1}) & = & m_{\ell}\overline{u(p_{2})}(C_{1}P_{R}+C_{2}P_{L})u(p_{1})\\
 & = & m_{\ell}\frac{C_{1}+C_{2}}{2}\overline{u(p_{2})}u(p_{1})+m_{\ell}\frac{C_{1}-C_{2}}{2}\overline{u(p_{2})}\gamma^{5}u(p_{1}).
\end{eqnarray*}
With this detail being noted, the calculation is straightforward and
gives:
\begin{equation}
i{\cal M}_{W}\approx-i\frac{G_{F}m_{1}m_{\ell}}{8\sqrt{2}\pi^{2}}\overline{u(p_{2})}\left[\frac{3}{4}(y_{R}+y_{R}^{*})+\frac{1}{2}(y_{R}-y_{R}^{*})\gamma^{5}\right]u(p_{1}).\label{eq:m-104}
\end{equation}
The first and second terms in the square bracket give rise to the
loop-induced scalar and pseudo-scalar couplings respectively. The
former has been considered in Sec.~\ref{sec:Loop}. The latter and
the $i{\cal M}_{Z}$ amplitude in Eq.~(\ref{eq:m-105})  lead to the
following pseudo-scalar interaction:
\begin{equation}
{\cal L}\supset\lambda_{\phi\ell\ell}^{(W)}\phi\overline{\psi_{\ell}}i\gamma^{5}\psi_{\ell}+\lambda_{\phi\ell\ell}^{(Z)}\phi\overline{\psi_{\ell}}i\gamma^{5}\psi_{\ell},\label{eq:m-103}
\end{equation}
where 
\begin{equation}
\lambda_{\phi\ell\ell}^{(W)}=-\frac{G_{F}m_{1}m_{\ell}{\rm Im}(y_{R})}{8\sqrt{2}\pi^{2}},\ \lambda_{\phi\ell\ell}^{(Z)}=-\frac{g_{Z}^{(\ell)}m_{1}m_{\ell}{\rm Im}(y_{R})}{32\pi^{2}m_{Z}^{2}}.\label{eq:m-55-1}
\end{equation}

\section{Generalization to three flavors\label{sec:3nu-app}}

In this appendix, we present the detailed three-flavor analysis. 

The $6\times6$ symmetric mass matrix can be diagonalized by a $6\times6$
unitary matrix $U$:
\begin{equation}
U^{T}M_{6\nu}U={\rm diag}(m_{1},\ m_{2},\ m_{3},\ \cdots,\ m_{6})\equiv M_{d}.\label{eq:m-62}
\end{equation}
The neutrino flavor basis and the mass basis are connected by
\begin{equation}
\left(\begin{array}{c}
\nu_{L}\\
\nu_{R}
\end{array}\right)=U\left(\begin{array}{c}
\nu_{1}\\
\vdots\\
\nu_{6}
\end{array}\right),\label{eq:m-58}
\end{equation}
where both $\nu_{L}$ and $\nu_{R}$ are $3\times1$ vectors. First,
we convert the gauge interaction $\frac{g}{\sqrt{2}}W_{\mu}^{-}\ell_{L}^{\dagger}\overline{\sigma}^{\mu}\nu_{L}$
from the flavor basis to the mass basis:
\begin{eqnarray}
{\cal L} & \supset & \frac{g}{\sqrt{2}}W_{\mu}^{-}(e_{L}^{\dagger},\ \mu_{L}^{\dagger},\ \tau_{L}^{\dagger})\overline{\sigma}^{\mu}\left(\begin{array}{cccccc}
1 &  &  & 0\\
 & 1 &  &  & 0\\
 &  & 1 &  &  & 0
\end{array}\right)U\left(\begin{array}{c}
\nu_{1}\\
\vdots\\
\nu_{6}
\end{array}\right).\label{eq:m-59}
\end{eqnarray}
This generalizes $g_{W}^{i}$ in Eq.~(\ref{eq:m-12}) from a $1\times2$
vector to a $3\times6$ matrix:
\begin{equation}
g_{W}^{\ell i}=\frac{g}{\sqrt{2}}U_{\ell i},\ \ (\ell=e,\ \mu,\ \tau,\ {\rm and}\ i=1\cdots6).\label{eq:m-61}
\end{equation}

Next, we perform a similar transformation for the Yukawa interactions
of $\nu_{R}$:
\begin{equation}
\frac{1}{2}\phi\nu_{R}^{T}Y_{R}^{0}\nu_{R}=\frac{\phi}{2}(\nu_{1},\ \cdots,\ \nu_{6})U^{T}\left(\begin{array}{cccc}
0\\
 & 0\\
 &  & 0\\
 &  &  & Y_{R}^{0}
\end{array}\right)U\left(\begin{array}{c}
\nu_{1}\\
\vdots\\
\nu_{6}
\end{array}\right)\equiv\frac{\phi}{2}(\nu_{1}\cdots\nu_{6})Y_{R}\left(\begin{array}{c}
\nu_{1}\\
\vdots\\
\nu_{6}
\end{array}\right),\label{eq:m-60}
\end{equation}
where the $6\times6$ matrix $Y_{R}$ is a generalization of the $2\times2$
$y_{R}^{ij}$ matrix in Eq.~(\ref{eq:m-11}). 

The generalization of $G^{ij}$ is quite straightforward. By replacing
$g_{W}^{i}$ and $y_{R}^{ij}$ in Eq.~(\ref{eq:m-48}) with $g_{W}^{\ell i}$
and $Y_{R}^{ij}$, we get:
\begin{equation}
G^{ij}=g_{W}^{\ell i*}g_{W}^{\ell j}(m_{j}Y_{R}^{ij}+m_{i}Y_{R}^{ij*}).\label{eq:m-57}
\end{equation}
As for $F_{1}$ and $F_{2}$, the expressions in Eqs.~(\ref{eq:F1})
and (\ref{eq:F2}) can be used directly except that now $i$ and $j$
run from 1 to 6 instead of 1 and 4.

With the generalized $G^{ij}$, $F_{1}$ and $F_{2}$, it is straightforward
to get the loop-induced effective Yukawa coupling in Eq.~(\ref{eq:m-63}).

Note that any constant terms in $F_{12}$ can be ignored because 
\begin{eqnarray}
 &  & \sum_{i,\thinspace j}U_{\ell i}^{*}U_{\ell j}\left(Y_{R}M_{d}+M_{d}Y_{R}^{\dagger}\right)_{ij}\nonumber \\
 & = & \left[\left(\begin{array}{cc}
1_{3\times3}\\
 & 0_{3\times3}
\end{array}\right)U^{*}U^{T}\left(\begin{array}{cc}
0_{3\times3}\\
 & Y_{R}^{0}
\end{array}\right)UM_{d}U^{T}\left(\begin{array}{cc}
1_{3\times3}\\
 & 0_{3\times3}
\end{array}\right)+{\rm h.c.}\right]_{\ell\ell}\nonumber \\
 & = & \left[\left(\begin{array}{cc}
1_{3\times3}\\
 & 0_{3\times3}
\end{array}\right)\left(\begin{array}{cc}
0_{3\times3}\\
 & Y_{R}^{0}
\end{array}\right)M_{6\nu}^{*}\left(\begin{array}{cc}
1_{3\times3}\\
 & 0_{3\times3}
\end{array}\right)+{\rm h.c.}\right]_{\ell\ell}\nonumber \\
 & = & 0,\label{eq:m-64}
\end{eqnarray}
which is similar to Eq.~(\ref{eq:m-85}).

Eq.~(\ref{eq:m-63}) applies for the most general $3\nu_{L}+3\nu_{R}$
scenario. Although its dependence on the PMNS matrix and light neutrino
masses is not manifest, each quantity in Eq.~(\ref{eq:m-63}) can
be readily evaluated using numerical methods. 

Below we would like to discuss a special case in which Eq.~(\ref{eq:m-63})
can be further simplified and expressed in the Casas-Ibarra parametrization~\cite{Casas:2001sr}.

If $M_{R}$ and $Y_{R}^{0}$ can be simultaneously diagonalized and
$m_{1,2,3}\ll m_{4,5,6}$, without loss of generality we can assume
$M_{R}$ and $Y_{R}^{0}$ are diagonal and $m_{D}$ can be expressed,
according to the Casas-Ibarra parametrization, as
\begin{equation}
m_{D}=iU_{L}^{*}\sqrt{m_{\nu}^{d}}R^{T}\sqrt{M_{R}},\label{eq:m-84}
\end{equation}
where $U_{L}$, $m_{\nu}^{d}$ and $R$ have been defined in Sec.~\ref{sec:3nu-1}. 

Then the full $6\times6$ mixing matrix $U$ can be approximately
decomposed as
\begin{equation}
U\approx\left(\begin{array}{cc}
U_{L}\\
 & 1_{3\times3}
\end{array}\right)\left(\begin{array}{cc}
1_{3\times3} & -iT\\
-iT^{\dagger} & 1_{3\times3}
\end{array}\right),\label{eq:m-88}
\end{equation}
where
\begin{equation}
T\equiv\sqrt{m_{\nu}^{d}}R^{\dagger}\sqrt{M_{R}^{-1}}.\label{eq:m-93}
\end{equation}
From Eqs.~(\ref{eq:m-84}) and (\ref{eq:m-93}), we have
\[
m_{D}M_{R}^{-1}=iU_{L}^{*}T^{*}.
\]

In the mass basis, the Yukawa coupling matrix $Y_{R}$, defined in
Eq.~(\ref{eq:m-60}), now reads
\begin{equation}
Y_{R}\approx\left(\begin{array}{cc}
{\cal O}(T^{2}) & -iT^{*}Y_{R}^{0}\\
-iY_{R}^{0}T^{\dagger} & Y_{R}^{0}
\end{array}\right).\label{eq:m-86}
\end{equation}

Next, we need to compute $F_{12}(m_{i},\ m_{j})$ in Eq.~(\ref{eq:m-63}).
Assuming $m_{4,5,6}\gg m_{W}\gg m_{1,2,3}$, the result is
\begin{equation}
F_{12}\approx\left(\begin{array}{cccccc}
4 & 4 & 4 & f_{4}-2 & f_{5}-2 & f_{6}-2\\
4 & 4 & 4 & f_{4}-2 & f_{5}-2 & f_{6}-2\\
4 & 4 & 4 & f_{4}-2 & f_{5}-2 & f_{6}-2\\
f_{4}-2 & f_{4}-2 & f_{4}-2 & f_{4}-8 & . & .\\
f_{5}-2 & f_{5}-2 & f_{5}-2 & . & f_{5}-8 & .\\
f_{6}-2 & f_{6}-2 & f_{6}-2 & . & . & f_{6}-8
\end{array}\right),\label{eq:m-75}
\end{equation}
where $f_{i}\equiv12\log\frac{m_{i}}{m_{W}}$ and ``$.$'' denotes
more complicated expressions which are irrelevant to our calculation.

By introducing the following matrix:
\begin{equation}
H_{ij}\equiv\left(Y_{R}M_{d}+M_{d}Y_{R}^{\dagger}\right)_{ij}F_{12}(m_{i},\ m_{j}),\label{eq:m-94}
\end{equation}
we can reformulate Eq.~(\ref{eq:m-63}) as
\begin{equation}
y_{\phi\ell\ell}=\frac{G_{F}m_{\ell}}{64\sqrt{2}\pi^{2}}\left[U^{*}HU^{T}\right]_{\ell\ell}.\label{eq:m-87}
\end{equation}

Combining Eqs.~(\ref{eq:m-86}) and (\ref{eq:m-75}), we obtain the
$H$ matrix:
\begin{equation}
H=\left(\begin{array}{cc}
{\cal O}(T^{4}) & -iT^{*}Y_{R}^{0}M_{R}D_{f}+{\cal O}(T^{3})\\
\left(-iT^{*}Y_{R}^{0}M_{R}D_{f}\right)^{\dagger}+{\cal O}(T^{3}) & Y_{R}^{0}M_{R}\left(D_{f}-6I_{3}\right)+{\rm h.c.}
\end{array}\right),\label{eq:m-89}
\end{equation}
where $D_{f}\equiv{\rm diag}(f_{4}-2,\ f_{5}-2,\ f_{6}-2)$ and $I_{3}$
is a $3\times3$ identity matrix.  Now supplying all the matrices
required by Eq.~(\ref{eq:m-87}), we obtain the results in Eqs.~(\ref{eq:m-90})
and (\ref{eq:m-92}).

\bibliographystyle{JHEP}
\bibliography{ref}

\providecommand{\href}[2]{#2}\begingroup\raggedright\begin{thebibliography}{10}

\bibitem{Asaka:2005pn}
T.~Asaka and M.~Shaposhnikov, {\it {The $\nu$MSM, dark matter and baryon
  asymmetry of the universe}},  {\em Phys. Lett. B} {\bf 620} (2005) 17--26,
  [\href{http://www.arxiv.org/abs/hep-ph/0505013}{{\tt hep-ph/0505013}}].

\bibitem{Asaka:2005an}
T.~Asaka, S.~Blanchet, and M.~Shaposhnikov, {\it {The nuMSM, dark matter and
  neutrino masses}},  {\em Phys. Lett. B} {\bf 631} (2005) 151--156,
  [\href{http://www.arxiv.org/abs/hep-ph/0503065}{{\tt hep-ph/0503065}}].

\bibitem{Alanne:2018brf}
T.~Alanne, T.~Hugle, M.~Platscher, and K.~Schmitz, {\it {Low-scale leptogenesis
  assisted by a real scalar singlet}},  {\em JCAP} {\bf 03} (2019) 037,
  [\href{http://www.arxiv.org/abs/1812.04421}{{\tt 1812.04421}}].

\bibitem{Minkowski:1977sc}
P.~Minkowski, {\it {$\mu \to e\gamma$ at a Rate of One Out of $10^{9}$ Muon
  Decays?}},  {\em Phys. Lett.} {\bf B67} (1977) 421--428.

\bibitem{yanagida1979proceedings}
T.~Yanagida, {\it {Proceedings of the Workshop on the Unified Theory and the
  Baryon Number in the Universe}},  {\em KEK Report 79-18} (1979) 95.

\bibitem{GellMann:1980vs}
M.~Gell-Mann, P.~Ramond, and R.~Slansky, {\it {Complex Spinors and Unified
  Theories}},  {\em Conf. Proc.} {\bf C790927} (1979) 315--321,
  [\href{http://www.arxiv.org/abs/1306.4669}{{\tt 1306.4669}}].

\bibitem{glashow1979future}
S.~Glashow, {\it {The future of elementary particle physics, NATO Adv}},  {\em
  Study Inst. Ser. B Phys} {\bf 59} (1979) 687.

\bibitem{mohapatra1980neutrino}
R.~Mohapatra and G.~Senjanovic, {\it Neutrino mass and spontaneous parity
  nonconservation},  {\em Phys. Rev. Lett.} {\bf 44} (1980), no.~14 912--915.

\bibitem{Turyshev:2006gm}
S.~G. Turyshev and J.~G. Williams, {\it {Space-based tests of gravity with
  laser ranging}},  {\em Int. J. Mod. Phys. D} {\bf 16} (2007) 2165--2179,
  [\href{http://www.arxiv.org/abs/gr-qc/0611095}{{\tt gr-qc/0611095}}].

\bibitem{Schlamminger:2007ht}
S.~Schlamminger, K.~Y. Choi, T.~A. Wagner, J.~H. Gundlach, and E.~G.
  Adelberger, {\it {Test of the equivalence principle using a rotating torsion
  balance}},  {\em Phys. Rev. Lett.} {\bf 100} (2008) 041101,
  [\href{http://www.arxiv.org/abs/0712.0607}{{\tt 0712.0607}}].

\bibitem{Heckel:2008hw}
B.~R. Heckel, E.~Adelberger, C.~Cramer, T.~Cook, S.~Schlamminger, and
  U.~Schmidt, {\it {Preferred-Frame and CP-Violation Tests with Polarized
  Electrons}},  {\em Phys. Rev. D} {\bf 78} (2008) 092006,
  [\href{http://www.arxiv.org/abs/0808.2673}{{\tt 0808.2673}}].

\bibitem{TheLIGOScientific:2017qsa}
{\bf LIGO Scientific, Virgo} {\bf Collaboration}, B.~Abbott {\em et~al.}, {\it
  {GW170817: Observation of Gravitational Waves from a Binary Neutron Star
  Inspiral}},  {\em Phys. Rev. Lett.} {\bf 119} (2017), no.~16 161101,
  [\href{http://www.arxiv.org/abs/1710.05832}{{\tt 1710.05832}}].

\bibitem{Abbott:2020khf}
{\bf LIGO Scientific, Virgo} {\bf Collaboration}, R.~Abbott {\em et~al.}, {\it
  {GW190814: Gravitational Waves from the Coalescence of a 23 Solar Mass Black
  Hole with a 2.6 Solar Mass Compact Object}},  {\em Astrophys. J.} {\bf 896}
  (2020), no.~2 L44, [\href{http://www.arxiv.org/abs/2006.12611}{{\tt
  2006.12611}}].

\bibitem{Croon:2017zcu}
D.~Croon, A.~E. Nelson, C.~Sun, D.~G.~E. Walker, and Z.-Z. Xianyu, {\it
  {Hidden-Sector Spectroscopy with Gravitational Waves from Binary Neutron
  Stars}},  {\em Astrophys. J. Lett.} {\bf 858} (2018), no.~1 L2,
  [\href{http://www.arxiv.org/abs/1711.02096}{{\tt 1711.02096}}].

\bibitem{Baryakhtar:2017ngi}
M.~Baryakhtar, R.~Lasenby, and M.~Teo, {\it {Black Hole Superradiance
  Signatures of Ultralight Vectors}},  {\em Phys. Rev.} {\bf D96} (2017), no.~3
  035019, [\href{http://www.arxiv.org/abs/1704.05081}{{\tt 1704.05081}}].

\bibitem{Sagunski:2017nzb}
L.~Sagunski, J.~Zhang, M.~C. Johnson, L.~Lehner, M.~Sakellariadou, S.~L.
  Liebling, C.~Palenzuela, and D.~Neilsen, {\it {Neutron star mergers as a
  probe of modifications of general relativity with finite-range scalar
  forces}},  {\em Phys. Rev. D} {\bf 97} (2018), no.~6 064016,
  [\href{http://www.arxiv.org/abs/1709.06634}{{\tt 1709.06634}}].

\bibitem{Hook:2017psm}
A.~Hook and J.~Huang, {\it {Probing axions with neutron star inspirals and
  other stellar processes}},  {\em JHEP} {\bf 06} (2018) 036,
  [\href{http://www.arxiv.org/abs/1708.08464}{{\tt 1708.08464}}].

\bibitem{Huang:2018pbu}
J.~Huang, M.~C. Johnson, L.~Sagunski, M.~Sakellariadou, and J.~Zhang, {\it
  {Prospects for axion searches with Advanced LIGO through binary mergers}},
  {\em Phys. Rev. D} {\bf 99} (2019), no.~6 063013,
  [\href{http://www.arxiv.org/abs/1807.02133}{{\tt 1807.02133}}].

\bibitem{Kopp:2018jom}
J.~Kopp, R.~Laha, T.~Opferkuch, and W.~Shepherd, {\it {Cuckoo's eggs in neutron
  stars: can LIGO hear chirps from the dark sector?}},  {\em JHEP} {\bf 11}
  (2018) 096, [\href{http://www.arxiv.org/abs/1807.02527}{{\tt 1807.02527}}].

\bibitem{Alexander:2018qzg}
S.~Alexander, E.~McDonough, R.~Sims, and N.~Yunes, {\it {Hidden-Sector
  Modifications to Gravitational Waves From Binary Inspirals}},  {\em Class.
  Quant. Grav.} {\bf 35} (2018), no.~23 235012,
  [\href{http://www.arxiv.org/abs/1808.05286}{{\tt 1808.05286}}].

\bibitem{Choi:2018axi}
H.~G. Choi and S.~Jung, {\it {New probe of dark matter-induced fifth force with
  neutron star inspirals}},  {\em Phys. Rev. D} {\bf 99} (2019), no.~1 015013,
  [\href{http://www.arxiv.org/abs/1810.01421}{{\tt 1810.01421}}].

\bibitem{Fabbrichesi:2019ema}
M.~Fabbrichesi and A.~Urbano, {\it {Charged neutron stars and observational
  tests of a dark force weaker than gravity}},  {\em JCAP} {\bf 06} (2020) 007,
  [\href{http://www.arxiv.org/abs/1902.07914}{{\tt 1902.07914}}].

\bibitem{Seymour:2019tir}
B.~C. Seymour and K.~Yagi, {\it {Probing Massive Scalar Fields from a Pulsar in
  a Stellar Triple System}},  \href{http://www.arxiv.org/abs/1908.03353}{{\tt
  1908.03353}}.

\bibitem{Dror:2019uea}
J.~A. Dror, R.~Laha, and T.~Opferkuch, {\it {Probing Muonic Forces with Neutron
  Stars Binaries}},  \href{http://www.arxiv.org/abs/1909.12845}{{\tt
  1909.12845}}.

\bibitem{Chikashige:1980ui}
Y.~Chikashige, R.~N. Mohapatra, and R.~D. Peccei, {\it {Are There Real
  Goldstone Bosons Associated with Broken Lepton Number?}},  {\em Phys. Lett.}
  {\bf 98B} (1981) 265--268.

\bibitem{Garcia-Cely:2017oco}
C.~Garcia-Cely and J.~Heeck, {\it {Neutrino Lines from Majoron Dark Matter}},
  {\em JHEP} {\bf 05} (2017) 102,
  [\href{http://www.arxiv.org/abs/1701.07209}{{\tt 1701.07209}}].

\bibitem{Heeck:2019guh}
J.~Heeck and H.~H. Patel, {\it {Majoron at two loops}},  {\em Phys. Rev. D}
  {\bf 100} (2019), no.~9 095015,
  [\href{http://www.arxiv.org/abs/1909.02029}{{\tt 1909.02029}}].

\bibitem{Joshipura:2003jh}
A.~S. Joshipura and S.~Mohanty, {\it {Constraints on flavor dependent long
  range forces from atmospheric neutrino observations at super-Kamiokande}},
  {\em Phys. Lett.} {\bf B584} (2004) 103--108,
  [\href{http://www.arxiv.org/abs/hep-ph/0310210}{{\tt hep-ph/0310210}}].

\bibitem{Grifols:2003gy}
J.~A. Grifols and E.~Masso, {\it {Neutrino oscillations in the sun probe long
  range leptonic forces}},  {\em Phys. Lett.} {\bf B579} (2004) 123--126,
  [\href{http://www.arxiv.org/abs/hep-ph/0311141}{{\tt hep-ph/0311141}}].

\bibitem{Bandyopadhyay:2006uh}
A.~Bandyopadhyay, A.~Dighe, and A.~S. Joshipura, {\it {Constraints on
  flavor-dependent long range forces from solar neutrinos and KamLAND}},  {\em
  Phys. Rev.} {\bf D75} (2007) 093005,
  [\href{http://www.arxiv.org/abs/hep-ph/0610263}{{\tt hep-ph/0610263}}].

\bibitem{GonzalezGarcia:2006vp}
M.~C. Gonzalez-Garcia, P.~C. de~Holanda, E.~Masso, and R.~Zukanovich~Funchal,
  {\it {Probing long-range leptonic forces with solar and reactor neutrinos}},
  {\em JCAP} {\bf 0701} (2007) 005,
  [\href{http://www.arxiv.org/abs/hep-ph/0609094}{{\tt hep-ph/0609094}}].

\bibitem{Nelson:2007yq}
A.~E. Nelson and J.~Walsh, {\it {Short Baseline Neutrino Oscillations and a New
  Light Gauge Boson}},  {\em Phys. Rev.} {\bf D77} (2008) 033001,
  [\href{http://www.arxiv.org/abs/0711.1363}{{\tt 0711.1363}}].

\bibitem{GonzalezGarcia:2008wk}
M.~C. Gonzalez-Garcia, P.~C. de~Holanda, and R.~Zukanovich~Funchal, {\it
  {Constraints from Solar and Reactor Neutrinos on Unparticle Long-Range
  Forces}},  {\em JCAP} {\bf 0806} (2008) 019,
  [\href{http://www.arxiv.org/abs/0803.1180}{{\tt 0803.1180}}].

\bibitem{Samanta:2010zh}
A.~Samanta, {\it {Long-range Forces : Atmospheric Neutrino Oscillation at a
  magnetized Detector}},  {\em JCAP} {\bf 1109} (2011) 010,
  [\href{http://www.arxiv.org/abs/1001.5344}{{\tt 1001.5344}}].

\bibitem{Heeck:2010pg}
J.~Heeck and W.~Rodejohann, {\it {Gauged $L_{\mu} - L_{\tau}$ and different
  Muon Neutrino and Anti-Neutrino Oscillations: MINOS and beyond}},  {\em J.
  Phys. G} {\bf 38} (2011) 085005,
  [\href{http://www.arxiv.org/abs/1007.2655}{{\tt 1007.2655}}].

\bibitem{Davoudiasl:2011sz}
H.~Davoudiasl, H.-S. Lee, and W.~J. Marciano, {\it {Long-Range Lepton Flavor
  Interactions and Neutrino Oscillations}},  {\em Phys. Rev.} {\bf D84} (2011)
  013009, [\href{http://www.arxiv.org/abs/1102.5352}{{\tt 1102.5352}}].

\bibitem{Heeck:2014zfa}
J.~Heeck, {\it {Unbroken B -- L symmetry}},  {\em Phys. Lett. B} {\bf 739}
  (2014) 256--262, [\href{http://www.arxiv.org/abs/1408.6845}{{\tt
  1408.6845}}].

\bibitem{Chatterjee:2015gta}
S.~S. Chatterjee, A.~Dasgupta, and S.~K. Agarwalla, {\it {Exploring
  Flavor-Dependent Long-Range Forces in Long-Baseline Neutrino Oscillation
  Experiments}},  {\em JHEP} {\bf 12} (2015) 167,
  [\href{http://www.arxiv.org/abs/1509.03517}{{\tt 1509.03517}}].

\bibitem{Bustamante:2018mzu}
M.~Bustamante and S.~K. Agarwalla, {\it {Universe's Worth of Electrons to Probe
  Long-Range Interactions of High-Energy Astrophysical Neutrinos}},  {\em Phys.
  Rev. Lett.} {\bf 122} (2019), no.~6 061103,
  [\href{http://www.arxiv.org/abs/1808.02042}{{\tt 1808.02042}}].

\bibitem{Khatun:2018lzs}
A.~Khatun, T.~Thakore, and S.~Kumar~Agarwalla, {\it {Can INO be Sensitive to
  Flavor-Dependent Long-Range Forces?}},  {\em JHEP} {\bf 04} (2018) 023,
  [\href{http://www.arxiv.org/abs/1801.00949}{{\tt 1801.00949}}].

\bibitem{Wise:2018rnb}
M.~B. Wise and Y.~Zhang, {\it {Lepton Flavorful Fifth Force and Depth-dependent
  Neutrino Matter Interactions}},  {\em JHEP} {\bf 06} (2018) 053,
  [\href{http://www.arxiv.org/abs/1803.00591}{{\tt 1803.00591}}].

\bibitem{Krnjaic:2017zlz}
G.~Krnjaic, P.~A.~N. Machado, and L.~Necib, {\it {Distorted neutrino
  oscillations from time varying cosmic fields}},  {\em Phys. Rev.} {\bf D97}
  (2018), no.~7 075017, [\href{http://www.arxiv.org/abs/1705.06740}{{\tt
  1705.06740}}].

\bibitem{Berlin:2016woy}
A.~Berlin, {\it {Neutrino Oscillations as a Probe of Light Scalar Dark
  Matter}},  {\em Phys. Rev. Lett.} {\bf 117} (2016), no.~23 231801,
  [\href{http://www.arxiv.org/abs/1608.01307}{{\tt 1608.01307}}].

\bibitem{Brdar:2017kbt}
V.~Brdar, J.~Kopp, J.~Liu, P.~Prass, and X.-P. Wang, {\it {Fuzzy dark matter
  and nonstandard neutrino interactions}},  {\em Phys. Rev.} {\bf D97} (2018),
  no.~4 043001, [\href{http://www.arxiv.org/abs/1705.09455}{{\tt 1705.09455}}].

\bibitem{Smirnov:2019cae}
A.~Y. Smirnov and X.-J. Xu, {\it {Wolfenstein potentials for neutrinos induced
  by ultra-light mediators}},  {\em JHEP} {\bf 12} (2019) 046,
  [\href{http://www.arxiv.org/abs/1909.07505}{{\tt 1909.07505}}].

\bibitem{Babu:2019iml}
K.~Babu, G.~Chauhan, and P.~Bhupal~Dev, {\it {Neutrino Non-Standard
  Interactions via Light Scalars in the Earth, Sun, Supernovae and the Early
  Universe}},  {\em Phys. Rev. D} {\bf 101} (2020), no.~9 095029,
  [\href{http://www.arxiv.org/abs/1912.13488}{{\tt 1912.13488}}].

\bibitem{Dreiner:2008tw}
H.~K. Dreiner, H.~E. Haber, and S.~P. Martin, {\it {Two-component spinor
  techniques and Feynman rules for quantum field theory and supersymmetry}},
  {\em Phys. Rept.} {\bf 494} (2010) 1--196,
  [\href{http://www.arxiv.org/abs/0812.1594}{{\tt 0812.1594}}].

\bibitem{Wilczek:1982rv}
F.~Wilczek, {\it {Axions and Family Symmetry Breaking}},  {\em Phys. Rev.
  Lett.} {\bf 49} (1982) 1549--1552.

\bibitem{Moody:1984ba}
J.~E. Moody and F.~Wilczek, {\it {NEW MACROSCOPIC FORCES?}},  {\em Phys. Rev.}
  {\bf D30} (1984) 130.

\bibitem{Patel:2015tea}
H.~H. Patel, {\it {Package-X: A Mathematica package for the analytic
  calculation of one-loop integrals}},  {\em Comput. Phys. Commun.} {\bf 197}
  (2015) 276--290, [\href{http://www.arxiv.org/abs/1503.01469}{{\tt
  1503.01469}}].

\bibitem{Smirnov:2018luj}
A.~Y. Smirnov and X.-J. Xu, {\it {Neutrino mixing in SO(10) GUTs with a
  non-Abelian flavor symmetry in the hidden sector}},  {\em Phys. Rev.} {\bf
  D97} (2018), no.~9 095030, [\href{http://www.arxiv.org/abs/1803.07933}{{\tt
  1803.07933}}].

\bibitem{Rodejohann:2017lre}
W.~Rodejohann and X.-J. Xu, {\it {Trimaximal $\mu$-$\tau$ reflection
  symmetry}},  {\em Phys. Rev.} {\bf D96} (2017), no.~5 055039,
  [\href{http://www.arxiv.org/abs/1705.02027}{{\tt 1705.02027}}].

\bibitem{Rodejohann:2015hka}
W.~Rodejohann and X.-J. Xu, {\it {A left-right symmetric flavor symmetry
  model}},  {\em Eur. Phys. J.} {\bf C76} (2016), no.~3 138,
  [\href{http://www.arxiv.org/abs/1509.03265}{{\tt 1509.03265}}].

\bibitem{Casas:2001sr}
J.~Casas and A.~Ibarra, {\it {Oscillating neutrinos and $\mu \to e, \gamma$}},
  {\em Nucl. Phys. B} {\bf 618} (2001) 171--204,
  [\href{http://www.arxiv.org/abs/hep-ph/0103065}{{\tt hep-ph/0103065}}].

\bibitem{Pearson:2018tkr}
J.~Pearson, N.~Chamel, A.~Potekhin, A.~Fantina, C.~Ducoin, A.~Dutta, and
  S.~Goriely, {\it {Unified equations of state for cold non-accreting neutron
  stars with Brussels--Montreal functionals -- I. Role of symmetry energy}},
  {\em Mon. Not. Roy. Astron. Soc.} {\bf 481} (2018), no.~3 2994--3026,
  [\href{http://www.arxiv.org/abs/1903.04981}{{\tt 1903.04981}}]. [Erratum:
  Mon.Not.Roy.Astron.Soc. 486, 768 (2019)].

\bibitem{Wagner:2012ui}
T.~Wagner, S.~Schlamminger, J.~Gundlach, and E.~Adelberger, {\it
  {Torsion-balance tests of the weak equivalence principle}},  {\em Class.
  Quant. Grav.} {\bf 29} (2012) 184002,
  [\href{http://www.arxiv.org/abs/1207.2442}{{\tt 1207.2442}}].

\bibitem{Adelberger:2009zz}
E.~Adelberger, J.~Gundlach, B.~Heckel, S.~Hoedl, and S.~Schlamminger, {\it
  {Torsion balance experiments: A low-energy frontier of particle physics}},
  {\em Prog. Part. Nucl. Phys.} {\bf 62} (2009) 102--134.

\bibitem{Adelberger:2006dh}
E.~G. Adelberger, B.~R. Heckel, S.~A. Hoedl, C.~D. Hoyle, D.~J. Kapner, and
  A.~Upadhye, {\it {Particle Physics Implications of a Recent Test of the
  Gravitational Inverse Sqaure Law}},  {\em Phys. Rev. Lett.} {\bf 98} (2007)
  131104, [\href{http://www.arxiv.org/abs/hep-ph/0611223}{{\tt
  hep-ph/0611223}}].

\bibitem{Bordag:2001qi}
M.~Bordag, U.~Mohideen, and V.~Mostepanenko, {\it {New developments in the
  Casimir effect}},  {\em Phys. Rept.} {\bf 353} (2001) 1--205,
  [\href{http://www.arxiv.org/abs/quant-ph/0106045}{{\tt quant-ph/0106045}}].

\bibitem{Davidson:2000hf}
S.~Davidson, S.~Hannestad, and G.~Raffelt, {\it {Updated bounds on millicharged
  particles}},  {\em JHEP} {\bf 05} (2000) 003,
  [\href{http://www.arxiv.org/abs/hep-ph/0001179}{{\tt hep-ph/0001179}}].

\bibitem{Redondo:2013lna}
J.~Redondo and G.~Raffelt, {\it {Solar constraints on hidden photons
  re-visited}},  {\em JCAP} {\bf 08} (2013) 034,
  [\href{http://www.arxiv.org/abs/1305.2920}{{\tt 1305.2920}}].

\bibitem{Boehm:2012gr}
C.~Boehm, M.~J. Dolan, and C.~McCabe, {\it {Increasing Neff with particles in
  thermal equilibrium with neutrinos}},  {\em JCAP} {\bf 1212} (2012) 027,
  [\href{http://www.arxiv.org/abs/1207.0497}{{\tt 1207.0497}}].

\bibitem{Kamada:2015era}
A.~Kamada and H.-B. Yu, {\it {Coherent Propagation of PeV Neutrinos and the Dip
  in the Neutrino Spectrum at IceCube}},  {\em Phys. Rev.} {\bf D92} (2015),
  no.~11 113004, [\href{http://www.arxiv.org/abs/1504.00711}{{\tt
  1504.00711}}].

\bibitem{Huang:2017egl}
G.-y. Huang, T.~Ohlsson, and S.~Zhou, {\it {Observational Constraints on Secret
  Neutrino Interactions from Big Bang Nucleosynthesis}},  {\em Phys. Rev.} {\bf
  D97} (2018), no.~7 075009, [\href{http://www.arxiv.org/abs/1712.04792}{{\tt
  1712.04792}}].

\bibitem{Kamada:2018zxi}
A.~Kamada, K.~Kaneta, K.~Yanagi, and H.-B. Yu, {\it {Self-interacting dark
  matter and muon $g-2$ in a gauged U$(1)_{L_{\mu} - L_{\tau}}$ model}},  {\em
  JHEP} {\bf 06} (2018) 117, [\href{http://www.arxiv.org/abs/1805.00651}{{\tt
  1805.00651}}].

\bibitem{Luo:2020sho}
X.~Luo, W.~Rodejohann, and X.-J. Xu, {\it {Dirac neutrinos and $N_{{\rm
  eff}}$}},  {\em JCAP} {\bf 06} (2020) 058,
  [\href{http://www.arxiv.org/abs/2005.01629}{{\tt 2005.01629}}].

\bibitem{Choi:1987sd}
K.~Choi, C.~W. Kim, J.~Kim, and W.~P. Lam, {\it {Constraints on the Majoron
  Interactions From the Supernova {SN1987A}}},  {\em Phys. Rev.} {\bf D37}
  (1988) 3225.

\bibitem{Choi:1989hi}
K.~Choi and A.~Santamaria, {\it {Majorons and Supernova Cooling}},  {\em Phys.
  Rev.} {\bf D42} (1990) 293--306.

\bibitem{Kachelriess:2000qc}
M.~Kachelriess, R.~Tomas, and J.~W.~F. Valle, {\it {Supernova bounds on Majoron
  emitting decays of light neutrinos}},  {\em Phys. Rev.} {\bf D62} (2000)
  023004, [\href{http://www.arxiv.org/abs/hep-ph/0001039}{{\tt
  hep-ph/0001039}}].

\bibitem{Hannestad:2002ff}
S.~Hannestad, P.~Keranen, and F.~Sannino, {\it {A Supernova constraint on bulk
  Majorons}},  {\em Phys. Rev.} {\bf D66} (2002) 045002,
  [\href{http://www.arxiv.org/abs/hep-ph/0204231}{{\tt hep-ph/0204231}}].

\bibitem{Farzan:2002wx}
Y.~Farzan, {\it {Bounds on the coupling of the Majoron to light neutrinos from
  supernova cooling}},  {\em Phys. Rev.} {\bf D67} (2003) 073015,
  [\href{http://www.arxiv.org/abs/hep-ph/0211375}{{\tt hep-ph/0211375}}].

\bibitem{Dent:2012mx}
J.~B. Dent, F.~Ferrer, and L.~M. Krauss, {\it {Constraints on Light Hidden
  Sector Gauge Bosons from Supernova Cooling}},
  \href{http://www.arxiv.org/abs/1201.2683}{{\tt 1201.2683}}.

\bibitem{Dreiner:2013mua}
H.~K. Dreiner, J.-F. Fortin, C.~Hanhart, and L.~Ubaldi, {\it {Supernova
  constraints on MeV dark sectors from $e^+e^-$ annihilations}},  {\em Phys.
  Rev. D} {\bf 89} (2014), no.~10 105015,
  [\href{http://www.arxiv.org/abs/1310.3826}{{\tt 1310.3826}}].

\bibitem{Bilmis:2015lja}
S.~Bilmis, I.~Turan, T.~Aliev, M.~Deniz, L.~Singh, and H.~Wong, {\it
  {Constraints on Dark Photon from Neutrino-Electron Scattering Experiments}},
  {\em Phys. Rev. D} {\bf 92} (2015), no.~3 033009,
  [\href{http://www.arxiv.org/abs/1502.07763}{{\tt 1502.07763}}].

\bibitem{Lindner:2016wff}
M.~Lindner, W.~Rodejohann, and X.-J. Xu, {\it {Coherent Neutrino-Nucleus
  Scattering and new Neutrino Interactions}},  {\em JHEP} {\bf 03} (2017) 097,
  [\href{http://www.arxiv.org/abs/1612.04150}{{\tt 1612.04150}}].

\bibitem{Farzan:2018gtr}
Y.~Farzan, M.~Lindner, W.~Rodejohann, and X.-J. Xu, {\it {Probing neutrino
  coupling to a light scalar with coherent neutrino scattering}},  {\em JHEP}
  {\bf 05} (2018) 066, [\href{http://www.arxiv.org/abs/1802.05171}{{\tt
  1802.05171}}].

\bibitem{Lindner:2018kjo}
M.~Lindner, F.~S. Queiroz, W.~Rodejohann, and X.-J. Xu, {\it {Neutrino-electron
  scattering: general constraints on $Z'$ and dark photon models}},  {\em JHEP}
  {\bf 05} (2018) 098, [\href{http://www.arxiv.org/abs/1803.00060}{{\tt
  1803.00060}}].

\bibitem{Khan:2019jvr}
A.~N. Khan, W.~Rodejohann, and X.-J. Xu, {\it {Borexino and general neutrino
  interactions}},  {\em Phys. Rev. D} {\bf 101} (2020), no.~5 055047,
  [\href{http://www.arxiv.org/abs/1906.12102}{{\tt 1906.12102}}].

\bibitem{Link:2019pbm}
J.~M. Link and X.-J. Xu, {\it {Searching for BSM neutrino interactions in dark
  matter detectors}},  {\em JHEP} {\bf 08} (2019) 004,
  [\href{http://www.arxiv.org/abs/1903.09891}{{\tt 1903.09891}}].

\bibitem{Poddar:2019wvu}
T.~Kumar~Poddar, S.~Mohanty, and S.~Jana, {\it {Vector gauge boson radiation
  from compact binary systems in a gauged $L_\mu-L_\tau$ scenario}},  {\em
  Phys. Rev. D} {\bf 100} (2019), no.~12 123023,
  [\href{http://www.arxiv.org/abs/1908.09732}{{\tt 1908.09732}}].

\bibitem{Reitze:2019iox}
D.~Reitze {\em et~al.}, {\it {Cosmic Explorer: The U.S. Contribution to
  Gravitational-Wave Astronomy beyond LIGO}},  {\em Bull. Am. Astron. Soc.}
  {\bf 51} (7, 2019) 035, [\href{http://www.arxiv.org/abs/1907.04833}{{\tt
  1907.04833}}].

\bibitem{Glashow:1970gm}
S.~L. Glashow, J.~Iliopoulos, and L.~Maiani, {\it {Weak Interactions with
  Lepton-Hadron Symmetry}},  {\em Phys. Rev.} {\bf D2} (1970) 1285--1292.

\end{thebibliography}\endgroup

\end{document}